\documentclass[11pt]{article}
\pdfoutput=1
\usepackage{authblk} 
\usepackage{cite} 
\usepackage{graphicx}
\usepackage{amsmath,amsfonts,amssymb}
\usepackage[margin=1in]{geometry}
\usepackage{array}
\usepackage{blkarray}
\usepackage{caption}
\usepackage{tikz}
\usepackage{listings}
\usetikzlibrary{snakes}
\usetikzlibrary{plotmarks}

\newcommand{\Rii}{
\begin{tikzpicture}[scale=0.5]
\draw [black, line width=0.2] (0,-1)  -- (1,0); 
\draw [black, line width=0.2] (0,1) -- (1,0); 
\draw [black, line width=0.2] (-1,0) -- (0,-1); 
\draw [black, line width=0.2] (-1,0) -- (0,1);  
\draw[blue, line width=0.3mm, rounded corners=7pt] (-0.5,-0.5) -- (-0.2,0.0) -- (-0.5,0.5);
\draw[blue, line width=0.3mm, rounded corners=7pt] (0.5,-0.5) -- (0.2,0.0) -- (0.5,0.5);
\draw[red, line width=0.3mm, rounded corners=7pt] (-0.4,-0.6) -- (0,0) -- (-0.4,0.6);
\draw[red, line width=0.3mm, rounded corners=7pt] (0.4,-0.6) -- (0,0) -- (0.4,0.6);
\end{tikzpicture}}
\newcommand{\Ref}{
\begin{tikzpicture}[scale=0.5]
\draw [black, line width=0.2] (0,-1)  -- (1,0); 
\draw [black, line width=0.2] (0,1) -- (1,0); 
\draw [black, line width=0.2] (-1,0) -- (0,-1); 
\draw [black, line width=0.2] (-1,0) -- (0,1);  
\draw[blue, line width=0.3mm, rounded corners=7pt] (-0.5,-0.5) -- (0.,0.0) -- (0.5,-0.5);
\draw[blue, line width=0.3mm, rounded corners=7pt] (-0.5,0.5) -- (0.,0.0) -- (0.5,0.5);
\draw[red, line width=0.3mm, rounded corners=7pt] (-0.4,-0.6) -- (-0,-0.2) -- (0.4,-0.6);
\draw[red, line width=0.3mm, rounded corners=7pt] (-0.4,0.6) -- (0,0.2) -- (0.4,0.6);
\end{tikzpicture}}
\newcommand{\Rei}{
\begin{tikzpicture}[scale=0.5]
\draw [black, line width=0.2] (0,-1)  -- (1,0); 
\draw [black, line width=0.2] (0,1) -- (1,0); 
\draw [black, line width=0.2] (-1,0) -- (0,-1); 
\draw [black, line width=0.2] (-1,0) -- (0,1);  
\draw[blue, line width=0.3mm, rounded corners=7pt] (-0.5,-0.5) -- (0.,0.0) -- (-0.5,0.5);
\draw[blue, line width=0.3mm, rounded corners=7pt] (0.5,-0.5) -- (0.,0.0) -- (0.5,0.5);
\draw[red, line width=0.3mm, rounded corners=7pt] (-0.4,-0.6) -- (-0,-0.2) -- (0.4,-0.6);
\draw[red, line width=0.3mm, rounded corners=7pt] (-0.4,0.6) -- (0,0.2) -- (0.4,0.6);
\end{tikzpicture}}
\newcommand{\Rif}{
\begin{tikzpicture}[scale=0.5]
\draw [black, line width=0.2] (0,-1)  -- (1,0); 
\draw [black, line width=0.2] (0,1) -- (1,0); 
\draw [black, line width=0.2] (-1,0) -- (0,-1); 
\draw [black, line width=0.2] (-1,0) -- (0,1);  
\draw[red, line width=0.3mm, rounded corners=7pt] (-0.5,-0.5) -- (0.,0.0) -- (-0.5,0.5);
\draw[red, line width=0.3mm, rounded corners=7pt] (0.5,-0.5) -- (0.,0.0) -- (0.5,0.5);
\draw[blue, line width=0.3mm, rounded corners=7pt] (-0.4,-0.6) -- (-0,-0.2) -- (0.4,-0.6);
\draw[blue, line width=0.3mm, rounded corners=7pt] (-0.4,0.6) -- (0,0.2) -- (0.4,0.6);
\end{tikzpicture}}
\newcommand{\iblue}{
\begin{tikzpicture}[scale=0.5]
\draw[blue, line width=0.3mm, rounded corners=7pt] (-0.5,-0.5) -- (-0,0) -- (-0.5,0.5);
\draw[blue, line width=0.3mm, rounded corners=7pt] (0.5,-0.5) -- (0,0) -- (0.5,0.5);
\end{tikzpicture}}
\newcommand{\ired}{
\begin{tikzpicture}[scale=0.5]
\draw[red, line width=0.3mm, rounded corners=7pt] (-0.5,-0.5) -- (-0,0) -- (-0.5,0.5);
\draw[red, line width=0.3mm, rounded corners=7pt] (0.5,-0.5) -- (0,0) -- (0.5,0.5);
\end{tikzpicture}}
\newcommand{\eblue}{
\begin{tikzpicture}[scale=0.5]
\draw[blue, line width=0.3mm, rounded corners=7pt] (-0.5,-0.5) -- (-0,0) -- (0.5,-0.5);
\draw[blue, line width=0.3mm, rounded corners=7pt] (-0.5,0.5) -- (0,0) -- (0.5,0.5);
\end{tikzpicture}}
\newcommand{\ered}{
\begin{tikzpicture}[scale=0.5]
\draw[red, line width=0.3mm, rounded corners=7pt] (-0.5,-0.5) -- (-0,0) -- (0.5,-0.5);
\draw[red, line width=0.3mm, rounded corners=7pt] (-0.5,0.5) -- (0,0) -- (0.5,0.5);
\end{tikzpicture}}

\newcommand{\RR}{\check{R}}

\title{Non compact continuum limit of two coupled Potts models}
\author[1,2]{\'Eric Vernier}
\author[1,3]{Jesper Lykke Jacobsen}
\author[2,4]{Hubert Saleur}
\affil[1]{LPTENS, \'Ecole Normale Sup\'erieure, 24 rue Lhomond, 75231 Paris, France}
\affil[2]{IPhT, CEA Saclay, 91191 Gif-sur-Yvette, France}
\affil[3]{Universit\'e Pierre et Marie Curie, 4 place Jussieu, 75252 Paris, France}
\affil[4]{USC Physics Department, Los Angeles CA 90089, USA}
\date{}
\begin{document}

\maketitle

\begin{abstract}

We study two $Q$-state Potts models coupled by the product of their
energy operators, in the regime $2 < Q \le 4$ where the coupling is
relevant. A particular choice of weights on the square lattice is shown to
be equivalent to the integrable $a_3^{(2)}$ vertex model. It corresponds
to a selfdual system of two antiferromagnetic Potts models, coupled
ferromagnetically. We derive the Bethe Ansatz equations and study them
numerically for two arbitrary twist angles. The continuum limit is shown to
involve two compact bosons and one non compact boson,
with discrete states emerging from the continuum at appropriate twists.
The non compact boson entails strong logarithmic corrections
to the finite-size behaviour of the scaling levels, the understanding
of which allows us to correct an earlier proposal for some of the critical
exponents. In particular, we infer
the full set of magnetic scaling dimensions (watermelon operators) of the Potts model.

\end{abstract}

\section{Introduction}

The two-dimensional Potts model pervades statistical physics and is a vivid
illustration of the strong ties between conformal field theory (CFT), integrability,
algebra, and probability theory. For $Q \in [0,4]$, and on
the square lattice, it exhibits integrable points corresponding to second-order phase transitions, in both the ferromagnetic \cite{Baxter73,BaxterBook} and the antiferromagnetic \cite{Baxter82,JS_AFPotts} regimes.

While the integrable
aspects emerge from transforming the Potts model into a vertex model, the conformal properties are best investigated by transforming it into a loop model \cite{LoopReview}.
The continuum limit of the ferromagnetic case is the well-studied compactified boson CFT. The antiferromagnetic case hase only been understood very recently: it corresponds to a compact boson coupled to another non compact boson \cite{IkhlefJS1,IkhlefJS2,IkhlefJS3,CanduIkhlef} and provides a statistical physics realisation of the Euclidean black hole sigma model \cite{BlackHoleCFT} which has been extensively studied in a string theory context \cite{Troost,RibSch}.

In the theory of disordered systems there is a strong motivation to study
the Potts model with quenched bond randomness. In a perturbative CFT approach
\cite{LudwigCardy87,DPP95} this corresponds to the replica limit ($N \to 0$) of a
system of $N$ Potts models coupled by the term
$g \int {\rm d}x \, \sum_{a \neq b} \varepsilon_a(x) \varepsilon_b(x)$ in the action, where $\varepsilon_a(x)$ denotes the local energy operator ($\Phi_{21}$ in Kac notation) of replica $a=1,2,\ldots,N$. This term is relevant in the renormalisation group (RG) sense for $Q>2$ --- an observation which provides the basis for the perturbative expansion around $Q=2$.
Unfortunately, apart from the perturbative CFT results, the progress on the random-bond Potts model has mainly been numerical \cite{Picco97,CJ97,JC98}.

Alternatively, one can study the coupled replicas for finite, integer $N \ge 2$. The perturbative CFT predicts a non-trivial fixed point $g_*$ for $N \ge 3$ for which a variety of critical exponents can be computed perturbatively, in agreement with numerical transfer matrix results and a duality analysis \cite{DJLP99,JJ00}. The case $N=2$ is
special, since all terms in the beta function, except the leading one, contain the factor
$(N-2)$. Accordingly the perturbative expressions for the critical exponents are singular at $N=2$.

Analytical progress has been limited, this far, to this case of $N=2$ coupled models.
For $Q=2$ this is known as the Ashkin-Teller model \cite{AT43}, which can be solved through a mapping to the eight-vertex model \cite{BaxterBook}. The coupling term is here exactly marginal and leads to a line of critical
points along which the critical exponents vary continuously.
For $Q>2$, and in the case where the Potts spins interact ferromagnetically, a field theoretical analysis reveals that the perturbation makes the model massive \cite{Vaysburd95}. This agrees with the duality analysis and transfer matrix computations \cite{DJLP99}, as in this case the two models couple strongly to form a single $Q^2$-state model which is non critical (since $Q^2 > 4$).

There however exists another, integrable case of two coupled Potts models, which was
found by Au-Yang and Perk \cite{Perk} by direct solution of the star-triangle equation;
see eq.~(2.17) in that paper. This line of integrable points
was further investigated by Martins and Nienhuis \cite{MartinsNienhuis} (it is called
solution 2 in their paper), who established the corresponding Bethe Ansatz
equations. They also showed that there are two regimes within the range $0 \le Q \le 4$,
each one corresponding to distinct critical behaviour. Martins and Nienhuis were
mainly interested in the case $Q=1$ which can be interpreted as a Lorentz lattice gas.
It belongs to the first regime, with $0 \le Q < 2$, throughout which the Potts interaction is
ferromagnetic and the two models decouple in the continuum limit. The second regime,
with $2 < Q \le 4$, was only treated briefly, and numerical evidence of an effective central
charge $c_{\rm eff} = 3$ was given.

For the remainder of this paper we shall focus on this second regime, $2 < Q \le 4$, for which
the Potts interaction is {\em antiferromagnetic} and
the energy-energy coupling between the two models is relevant by the perturbative
CFT analysis. Fendley and Jacobsen \cite{FJ08} presented a detailed analysis of this case,
based on the level-rank duality \cite{FK09} of the $SO(N)_k$ Birman-Wenzl-Murakami (BWM) algebras \cite{BWM}.
Parameterising $Q$ as
\begin{equation}
 \sqrt{Q} = 2 \cos \left( \frac{\pi}{k+2} \right) \,,
 \label{eq:loopweight}
\end{equation}
they found that these theories correspond \cite{FJ08} to the conformal coset
\begin{equation}
 \frac{SO(k)_3 \times SO(k)_1}{SO(k)_4} \approx
 \frac{SU(2)_k \times SU(2)_k}{SU(2)_{2k}}
\end{equation}
with central charge
\begin{equation}
 c = \frac{3 k^2}{(k+1)(k+2)} \,.
 \label{eq:cc}
\end{equation}

The purpose of this paper is to study further this integrable case of two coupled
antiferromagnetic Potts models. We shall see that the integrable
$\check{R}$ matrix is equivalent to that of $U_q(sl_4^{(2)})$ --- also known as
the $a_3^{(2)}$ model --- in the fundamental representation \cite{GalleasMartins,GalleasMartins04}. This allows us in particular to identify and study in details the corresponding Bethe Ansatz equations.%
\footnote{
The Bethe Ansatz equations already appeared in \cite{MartinsNienhuis}, but they were not
subjected to a systematic investigation in the range $2 < Q \le 4$.
}

On a more fundamental level, this equivalence places the two coupled Potts
models into the family of $a_n^{(2)}$ models whose first member we have recently investigated in detail \cite{VJS:a22}.
This $a_2^{(2)}$ model is related to the well-known O($n$) model on the square lattice \cite{Nienhuis}, and our study \cite{VJS:a22} concentrated on the so-called regime III, a model of dilute loops that contains a special point ($n \to 0$) which is a candidate for describing the theta-point collapse of polymers \cite{VJS:polymers}.
Using both analytical arguments and extensive numerical analysis we established that the $a_2^{(2)}$ model in regime III has a non compact continuum limit that turns out to be precisely the same as that of the antiferromagnetic Potts model \cite{JS_AFPotts,IkhlefJS1,IkhlefJS2,IkhlefJS3,CanduIkhlef}, namely that of the
$SL(2,\mathbb{R})_k/U(1)$ Euclidean black hole CFT \cite{BlackHoleCFT,Troost,RibSch}.

We show here that the $a_3^{(2)}$ model, relevant for describing two coupled Potts models, also has a non compact continuum limit, albeit now involving three rather than two bosons [cf.\ eq.~(\ref{eq:cc})]. The range $2 < Q \le 4$ in which the Potts models couple non-trivially corresponds precisely to the interesting regime III. Just like in the $a_2^{(2)}$ counterpart, the ``spectrum'' of critical exponents in the $a_3^{(2)}$ model contains both continuous and discrete states, with the discrete states emerging from --- and redisappearing into --- the continuum upon changing the twist. The twist is here controlled by two angles (rather than one in the $a_2^{(2)}$ case) that correspond to modifying the weights of the non contractible loops in each of the two Potts models. We provide the critical exponents of the magnetic-type operators, as functions of the two twists, and infer from those the scaling dimensions $x_{2n_1,2n_2}$ of the so-called watermelon operators in the Potts model, corresponding to the insertion of any given number $(2n_1,2n_2)$ of propagating ``through-lines'' in 
each of the two Potts models.

To keep the presentation light, the analysis given here is mainly based on analogies with the $a_2^{(2)}$ case and on an extensive numerical analysis of the Bethe Ansatz equations. A more formal treatment, that corroborates the present analysis, will appear elsewhere in the general $a_n^{(2)}$ context \cite{VJS:an2}.

The fact that the continuum limit is non compact implies that the finite-size free energies, from which the critical exponents are extracted in the usual way, often contain strong logarithmic corrections to scaling. Ref.~\cite{FJ08} made an attempt of conjecturing the first few watermelon exponents based on direct diagonalisation of the 
transfer matrix for sizes up to $L=16$ loop strands. Our present knowledge of the logarithmic corrections, combined with the ability to numerically solve the Bethe Ansatz equations for vastly larger sizes (typically $L \simeq 100$), obviously gives a much stronger handle on this problem. It is therefore hardly surprising that a few of the conjectures presented in \cite{FJ08} turn out to be wrong.

The plan of this paper is the following. In section \ref{section:models} we review the definition of the two coupled Potts models and their equivalence with the two-colour dense loop model studied in \cite{FJ08}. We then show how the latter can be reformulated in terms of a properly twisted $a_{3}^{(2)}$ model. The Bethe Ansatz study of the $a_3^{(2)}$ model is presented in section \ref{section:BAEa32}, allowing us to compute the conformal spectrum, which turns out to exhibit non compact features. These results are then applied to the calculation of the loop model's critical exponents in section \ref{section:confloop}. We give general formulae for the watermelon exponents, some of which differ significantly from the numerical estimations of \cite{FJ08}. Our findings are summarised and discussed in section~\ref{section:conclusion}.

\section{From two coupled Potts models to the $a_3^{(2)}$ vertex model}
\label{section:models}

We wish to study a system of two coupled Potts models described by the
Hamiltonian
\begin{equation}
 {\cal H} = - \sum_{\langle ij \rangle} \left[
 K(\delta_{\sigma_i,\sigma_j} + \delta_{\tau_i,\tau_j}) +
 L \delta_{\sigma_i,\sigma_j} \delta_{\tau_i,\tau_j} \right] \,,
 \label{eq:hamiltonian}
\end{equation}
where $\langle ij \rangle$ denotes the set of nearest neighbour sites (edges) on the square
lattice, and the Kronecker symbol $\delta_{x,y}$ equals $1$ if $x=y$, and $0$
otherwise. The spins $\sigma_i$ and $\tau_i$ of the first and second models take
the values $1,2,\ldots,Q$.
[The extension to two different models with $Q_1$ and $Q_2$ states
is interesting, but does not to our knowledge sustain an integrable formulation.]
Writing ${\cal H} = -\sum_{\langle ij \rangle} {\cal H}_{ij}$ the local Boltzmann
weight becomes
\begin{equation}
 W_{ij} \equiv \mathrm{e}^{-{\cal H}_{ij}} =
 1 + v (\delta_{\sigma_i,\sigma_j} + \delta_{\tau_i,\tau_j}) +
 (v^2 + w(1+v)^2) \delta_{\sigma_i,\sigma_j} \delta_{\tau_i,\tau_j} \,,
 \label{boltzmann}
\end{equation}
where we have defined $v = \mathrm{e}^K - 1$ and $w = \mathrm{e}^L - 1$.
The duality analysis \cite{DomanyRiedel,DJLP99,JJ00} shows that
selfduality is attained by setting the coefficient of the $\delta_{\sigma_i,\sigma_j} \delta_{\tau_i,\tau_j}$ term 
to $Q$, viz.
\begin{equation}
 w = \frac{Q - v^2}{(1+v)^2} \,.
 \label{selfdual}
\end{equation}

\subsection{Loop model and integrable $\check{R}$ matrix}

The partition function is obtained by expanding the product over $W_{ij}$
and summing over the spins,
\begin{equation}
 Z = \sum_{\{\sigma,\tau\}} \prod_{\langle ij \rangle} W_{ij} \,.
\end{equation}
It is convenient to associate a graphical representation with this expansion.
We first concentrate on just the first Potts model. For a horizontal edge $(ij)$ we
draw the edge or leave it empty
\begin{equation}
\begin{tikzpicture}[scale=0.5]
 \draw [black, line width=0.2] (0,-1)  -- (1,0); 
 \draw [black, line width=0.2] (0,1) -- (1,0); 
 \draw [black, line width=0.2] (-1,0) -- (0,-1); 
 \draw [black, line width=0.2] (-1,0) -- (0,1);  
 \draw [black,line width=1.0] (-1,0) -- (1,0);
 \draw (-1,0) node[left] {$\delta_{\sigma_i,\sigma_j} \equiv$};
\end{tikzpicture}
\qquad
\begin{tikzpicture}[scale=0.5]
 \draw [black, line width=0.2] (0,-1)  -- (1,0); 
 \draw [black, line width=0.2] (0,1) -- (1,0); 
 \draw [black, line width=0.2] (-1,0) -- (0,-1); 
 \draw [black, line width=0.2] (-1,0) -- (0,1);  
 \draw [black, dashed,line width=1.0] (-1,0) -- (1,0);
 \draw (-1,0) node[left] {$1 \equiv$};
\end{tikzpicture}
\end{equation}
depending on whether we take a term with or without the $\delta_{\sigma_i,\sigma_j}$ interaction. The set of drawn edges form a set of connected clusters, and the sum over $\{\sigma\}$ amounts to giving a weight $Q$ per cluster. Equivalently, we draw loops on the medial lattice \cite{BKW76,DJLP99}
\begin{equation}
\begin{tikzpicture}[scale=0.5]
 \draw [black, line width=0.2] (0,-1)  -- (1,0); 
 \draw [black, line width=0.2] (0,1) -- (1,0); 
 \draw [black, line width=0.2] (-1,0) -- (0,-1); 
 \draw [black, line width=0.2] (-1,0) -- (0,1);  
 \draw[red, line width=0.3mm, rounded corners=7pt] (-0.5,-0.5) -- (0.,0.0) -- (0.5,-0.5);
 \draw[red, line width=0.3mm, rounded corners=7pt] (-0.5,0.5) -- (0.,0.0) -- (0.5,0.5);
 \draw [black,line width=1.0] (-1,0) -- (1,0);
 \draw (-1,0) node[left] {$\delta_{\sigma_i,\sigma_j} \equiv$};
\end{tikzpicture}
\qquad
\begin{tikzpicture}[scale=0.5]
 \draw [black, line width=0.2] (0,-1)  -- (1,0); 
 \draw [black, line width=0.2] (0,1) -- (1,0); 
 \draw [black, line width=0.2] (-1,0) -- (0,-1); 
 \draw [black, line width=0.2] (-1,0) -- (0,1);  
 \draw[red, line width=0.3mm, rounded corners=7pt] (-0.4,-0.6) -- (0,0) -- (-0.4,0.6);
 \draw[red, line width=0.3mm, rounded corners=7pt] (0.4,-0.6) -- (0,0) -- (0.4,0.6);
 \draw [black, dashed,line width=1.0] (-1,0) -- (1,0);
 \draw (-1,0) node[left] {$1 \equiv$};
\end{tikzpicture}
\end{equation}
such that the loops bounce off the empty edges and cut through the occupied edges.
Using the Euler relation this provides a weight $n = \sqrt{Q}$ per closed loop, and at the
selfdual point (\ref{selfdual}) the local Boltzmann weight can be represented as
\begin{equation}
 W_{ij} = \raisebox{-0.4cm}{\Rii} + \lambda \left( \raisebox{-0.4cm}{\Rei} +
 \raisebox{-0.4cm}{\Rif} \right) + \raisebox{-0.4cm}{\Ref} \,,
\end{equation}
where we have represented the loops corresponding to the second Potts model
by a different colour and defined $\lambda = v / \sqrt{Q}$. Note that due to the
selfduality the weights are now invariant under a $90^\circ$ rotation, so we get
the same expression for horizontal and vertical edges. Accordingly, we have omitted
the graphical rendering of the edge itself, retaining only the loops.

It is this dense two-colour loop model that was studied in \cite{FJ08}.
The local Boltzmann weights define the corresponding $\RR$ matrix, so
we shall henceforth write
\begin{equation}
 \RR =  \left( \raisebox{-0.4cm}{\Rii} + \raisebox{-0.4cm}{\Ref} \right) +
 \lambda_c \left( \raisebox{-0.4cm}{\Rei} + \raisebox{-0.4cm}{\Rif} \right) \,,
 \label{RJF}
\end{equation}
where now $\lambda_c$ is the integrable choice of the coupling constant.
It is given by
\begin{equation}
 \lambda_c = \frac12 \left( -\sqrt{Q} + \sqrt{4-Q} \right) \,.
 \label{lambdacQ}
\end{equation}
It is convenient to parameterise the loop weight by
$n = 2 \cos \gamma$, with $\gamma = {\pi \over k+2}$, 
in agreement with (\ref{eq:loopweight}). The critical coupling is then
\begin{equation}
 \lambda_c = -\sqrt{2} \sin \left({\pi \over 4} {k-2 \over k+2}\right) \,,
 \label{cJF}
\end{equation}
and the central charge found from the level-rank duality argument
of \cite{FJ08} is given by (\ref{eq:cc}).

We can now interpret the value (\ref{cJF}) physically in terms of the
spin interaction $K$ and the coupling between models $L$ appearing
in the original Hamiltonian (\ref{eq:hamiltonian}). We are interested in
the regime $2 < Q \le 4$ where the two Potts models couple non trivially.
Within this regime, $K$ is real and negative provided that
$\lambda_c \ge -1/\sqrt{Q}$ --- that is $2 < Q \le 2 + \sqrt{2}$, or $2 < k \le 6$ ---
so the spins interact antiferromagnetically.%
\footnote{For $2 + \sqrt{2} < Q \le 4$, or $k > 6$,
the original Potts formulation (\ref{eq:hamiltonian}) is unphysical, corresponding
to complex $K$, but the two-colour loop formulation still makes sense.}
On the other hand,
\begin{equation}
 w = \frac{2 Q \sqrt{Q(4-Q)}}
 {\left( 2 - Q + \sqrt{Q(4-Q)} \right)^2}
\end{equation}
is real and non-negative for any $Q \in [0,4]$, and so is $L$.

For integer $k$ the coupled Potts models can also be formulated as an RSOS height model
whose weights can be brought into positive definite form \cite{FJ08} under the same condition,
that is, $2 < k \le 6$. 

It was observed in \cite{FJ08} that (\ref{RJF}) is the isotropic point of a more general, spectral parameter dependent, integrable $\RR$ matrix. 
Let us recall this construction (more details are provided in \cite{FJ08}).

Each loop colour (red or blue) is independently a representation of the 
Temperley-Lieb (TL) algebra \cite{TL}. Its generators satisfy the well-known
relations
\begin{eqnarray}
 e_{(i)} e_{(i)} &=& n e_{(i)} \,, \nonumber \\
 e_{(i)} e_{(i \pm 1)} e_{(i)} &=& e_{(i)} \,, \label{TLrelations} \\ 
 e_{(i)} e_{(j)} &=& e_{(j)} e_{(i)} \mbox{ if $|i-j| > 1$} \,. \nonumber
\end{eqnarray}
Omitting henceforth the site index, we make the following graphical
identification of the identity operator and TL generator in the two models:
\begin{eqnarray}
 i_1 = \raisebox{-0.2cm}{\iblue} & & \qquad e_1 = \raisebox{-0.2cm}{\eblue}  \\
 i_2 = \raisebox{-0.2cm}{\ired} & & \qquad e_2 = \raisebox{-0.2cm}{\ered} 
 \label{TLonecolour}
\end{eqnarray}
The integrable $\RR$ matrix (\ref{RJF}) then reads
\begin{equation}
 \RR = i_1 \otimes i_2 + e_1 \otimes e_2 + \lambda_c \left(e_1 \otimes i_2  +  i_1 \otimes e_2 \right) \,.
 \label{Re1e2}
\end{equation}
Setting now 
\begin{eqnarray}
I  &=& i_1 \otimes i_2 \,, \\
E &=& e_1 \otimes e_2 \,, \\
B &=& \left( q^{- {1 \over 2}}i_1 - q^{1 \over 2} e_1 \right) \otimes  \left( q^{- {1 \over 2}}i_2 - q^{1 \over 2} e_2 \right) \,,
\end{eqnarray}
where we defined $q=\mathrm{e}^{\mathrm{i}\gamma} = \mathrm{e}^{\mathrm{i}{\pi \over k+2}}$, it is straightforward to see that $I, E, B \equiv I^{(4,k)}, E^{(4,k)}, B^{(4,k)}$ are the generators of an $SO(4)_k$ Birman-Wenzl-Murakami (BWM) algebra \cite{BWM}.
The $SO(4)_k$ algebra is part of the more general family of $SO(N)_k$ BWM algebras ($N$ and $k$, both integers, are called respectively the {\it rank} and {\it level}), which enjoy the interesting property of {\it level-rank duality}\/: the generators of $SO(N)_k$ can be rewritten in terms of those of $SO(k)_N$, and vice-versa. 

The construction of integrable $\RR$ matrices based on BWM generators is well known \cite{Akutsu}. 
For $SO(k)_N$ one defines an integrable, spectral parameter dependent model as 
\begin{equation}
 \check{R}^{(N,k)} = \left[{N \over 2}-1 - u\right] \left[1-u\right]I^{(N,k)} +  \left[u\right] \left[2-{N \over 2}+u\right] E^{(N,k)} +  \left[{N \over 2}-1-u\right] \left[u\right]X^{(N,k)} \,,
\end{equation}
where $[x]\equiv {q^{x}-q^{-x} \over q - q^{-1}}$ with now $q=\mathrm{e}^{\mathrm{i}{\pi \over N+k-2}}$, and $X^{(N,k)}=q^{-1}I^{(N,k)} + q E^{(N,k)} - B^{(N,k)}$.

Following \cite{FJ08} we start from $\check{R}^{(k,4)}$, which can be rewritten in terms of the $SO(4)_k$ generators using level-rank duality. 
The result is, after proper normalization (see section 3.1 of \cite{FJ08} for details),
\begin{equation}
 \check{R}^{(k,4)} =  I 
 + {\sin\left( {\pi u \over 2+k} \right) \over  \sin\left(\pi { 1+u \over 2+k} \right) } \left( 2\cos\left({\pi  \over 2+k}\right) + {\cos\left(\pi {1+u-k \over k+2} \right) \over  \cos\left(\pi {2+u \over 2+k} \right)} \right) E 
 - {\sin\left( {\pi u \over 2+k} \right) \over  \sin\left(\pi { 1+u \over 2+k} \right) } X \,.
 \label{Rk4}
\end{equation}
At the isotropic point $u = {k\over4}-{1\over 2}$ this is exactly the decomposition (\ref{RJF}).

\subsection{Formulation as an integrable vertex model}

Having at hand an integrable, spectral parameter dependent $\RR$ matrix for describing the critical point (\ref{RJF}) allows us to look for a Bethe ansatz solution. 
To proceed, we need to find a representation of (\ref{Rk4}) which is purely algebraic, as the $\RR$ matrix of some vertex model. This is done following the lines of \cite{GalleasMartins}, where the $\RR$ matrix based on the generators of BWM algebras are rewritten as those of certain $q$-deformed Lie (super)algebras. 
In our case, we see that $\check{R}^{(k,4)}$ correspond to the integrable $\RR$ matrices associated with the superalgebras $U_q \left( sl(4+r | r)^{(2)} \right)$, with $r=0,1,\dots$, whose matrix expression in the tensor product of fundamental representations is given explicitly in \cite{GalleasMartins04}.

Restricting to the simplest of these representations, namely $r=0$, we therefore arrive at the conclusion that (\ref{Rk4}) is equivalent to the integrable $\RR$ matrix associated with $U_q \left(sl_4^{(2)}\right)$, which more commonly goes by the name of $a_3^{(2)}$ model. Since we will be concerned with this model from now on, it is worthwhile  recalling its definition explicitly.

Consider a system of horizontal length $L$, each site of which carries a space $\mathcal{V}\equiv \mathbb{C}^{4}$. The spectral parameter dependent row-to-row transfer matrix is written as a trace over an auxilliary space $\mathcal{A}\equiv \mathbb{C}^{4}$, namely
\begin{equation}
 T^{(L)}(\lambda) = \mbox{Tr}_{\mathcal{A}} \left( R_{\mathcal{A}1}(\lambda)\ldots R_{\mathcal{A}L}(\lambda) \tau_{\mathcal{A}} \right) \,,
 \label{T6GM}
\end{equation}
where the matrix $R_{\mathcal{A}i}$ acts on the tensor product of the auxilliary space with the $i$th vertical --- or quantum --- space, and as the identity on the others. It is related to the $\RR$ matrix by a permutation or spaces, $R_{\mathcal{A}i} = \mathcal{P}_{\mathcal{A}i} \RR_{\mathcal{A}i}$. Moreover, the twist operator $\tau_{\mathcal{A}}$ acts diagonally on $\mathcal{A}$ in a way that will be made explicit later, and the $\RR_{ab}$ matrix acting on two spaces $a,b$ can be decomposed in term of the $4 \times 4$ Weyl matrices in $a$ and $b$ as \cite{GalleasMartins04}
\begin{eqnarray}
\RR_{ab}(\lambda) &=& a(\lambda) \sum_{\stackrel{\alpha=1}{\alpha \neq \alpha'}}^{4} 
\hat{e}^{(a)}_{\alpha \alpha} \otimes \hat{e}^{(b)}_{\alpha \alpha}
+b (\lambda) \sum_{\stackrel{\alpha ,\beta=1}{\alpha \neq \beta,\alpha \neq \beta'}}^{4}
\hat{e}^{(a)}_{\beta \alpha} \otimes \hat{e}^{(b)}_{\alpha \beta}  \nonumber \\
&+& {\bar{c}} (\lambda) \sum_{\stackrel{\alpha ,\beta=1}{\alpha < \beta,\alpha \neq \beta'}}^{4} \hat{e}^{(a)}_{\alpha \alpha} \otimes \hat{e}^{(b)}_{\beta \beta}
+c (\lambda) \sum_{\stackrel{\alpha ,\beta=1}{\alpha > \beta,\alpha \neq \beta'}}^{4} \hat{e}^{(a)}_{\alpha \alpha} \otimes \hat{e}^{(b)}_{\beta \beta} \nonumber \\
&+& \sum_{\alpha ,\beta =1}^{4} d_{\alpha, \beta} (\lambda)
\hat{e}^{(a)}_{\alpha' \beta} \otimes \hat{e}^{(b)}_{\alpha \beta'} \,.
\label{RGM}
\end{eqnarray}
In the above formula every index $\alpha=1,\ldots,4$ corresponds to a conjugate index $\alpha' \equiv 5-\alpha$, and $\hat{e}^{(a)}_{\alpha \beta}$ (resp. $\hat{e}^{(b)}_{\alpha \beta}$) denotes the matrix acting on $a$ (resp. $b$) such that $\left(\hat{e}^{(a,b)}_{\alpha \beta}\right)_{\mu \nu} = \delta_{\alpha \mu} \delta_{\beta \nu}$. The Boltzmann weights  $a(\lambda)$,
$b(\lambda)$, $c(\lambda)$ and ${\bar{c}} (\lambda)$ are determined by
\begin{eqnarray}
\label{bw1}
a (\lambda) &=&(e^{2 \lambda} -\zeta)(e^{2 \lambda} -q^2) \,, \\
b (\lambda) &=&q(e^{2 \lambda} -1)(e^{2 \lambda} -\zeta) \,, \\
c (\lambda) &=&(1-q^2)(e^{2 \lambda} -\zeta) \,, \\
{\bar{c}} (\lambda) &=& e^{2 \lambda} c(\lambda) \,,
\end{eqnarray}
where $\zeta = -q^4$, whilst $d_{\alpha\beta}(\lambda)$ has the form
\begin{equation}
d_{\alpha, \beta} (\lambda) = \left \lbrace
\begin{array}{ll}
 q(e^{2 \lambda} -1)(e^{2 \lambda} -\zeta) +e^{2\lambda}(q^2 -1)(\zeta -1) &
 \mbox{for } \alpha=\beta=\beta' \,, \\
 (e^{2 \lambda} -1)\left[ (e^{2 \lambda} -\zeta)q^{2} +e^{2\lambda}(q^2 -1) \right] &
 \mbox{for } \alpha=\beta \neq \beta' \,, \\
 (q^{2 }-1)\left[ \zeta(e^{2 \lambda} -1) q^{t_{\alpha}-t_{\beta}} -\delta_{\alpha ,\beta'} (e^{2\lambda} -\zeta) \right] &
 \mbox{for } \alpha < \beta \,, \\
 (q^{2 }-1) e^{2 \lambda} \left[ (e^{2 \lambda} -1) q^{t_{\alpha}-t_{\beta}} -\delta_{\alpha ,\beta'} (e^{2\lambda} -\zeta) \right] &
 \mbox{for } \alpha > \beta \,, \\
 \end{array} \right.
\end{equation}
where $t_{\alpha}=-1,0,0,1$ for $\alpha=1,2,3,4$ respectively.
The exact identification of (\ref{Rk4}) with (\ref{RGM}) in fact involves some gauge changes, which we detail in the next section. 
It is important to notice that even though the use of level-rank duality for the BWM algebra is only defined for integer values of $k$, we are now left with a parametrization $\gamma = {\pi \over k+2}$ where $\gamma$, and therefore $k$, can vary continuously. 
More precisely we will consider in general $\gamma \in [0, {\pi \over 2}]$, because of the periodicity and the $\gamma \to {\pi - \gamma}$ symmetry of (\ref{RGM}).

The isotropic value of the spectral parameter $\lambda$ which recovers (\ref{RJF})--(\ref{lambdacQ})
is the following:
\begin{equation}
 \lambda_+ = \mathrm{i} \left( \gamma - {\pi \over 4} \right) \,.
 \label{lambdaisoRIII}
\end{equation}
Note that there is another value of the spectral parameter yielding an isotropic model, namely
\begin{equation}
 \lambda_- = \mathrm{i} \left( \gamma + {\pi \over 4} \right) \,.
 \label{lambdaisoRI}
\end{equation}
It corresponds to the other solution of $(\lambda_c)^2 + \lambda_c \sqrt{Q} + \frac12 Q = 1$, that is, replacing (\ref{lambdacQ}) by
\begin{equation}
 \lambda_c^{\pm} = \frac12 \left( -\sqrt{Q} \pm \sqrt{4-Q} \right) \,.
\end{equation}
The leading eigenvalues at one or another of these isotropic points do not correspond to the same eigenstates, and therefore define different regimes. We shall come back to this issue in section \ref{section:BAEa32}.

Note also that we only consider here periodic or twisted periodic boundary conditions. It would however also be interesting to study the case where the system has open boundary conditions in the horizontal directions. Integrable reflection matrices need to be introduced in this case, and we point out that one solution has been found in \cite{Malara,GalleasOpen}. This solution contains a free parameter, which is reminiscent of the situation for a single Potts model where the apperance
of an arbitrary constant of separation \cite{Doikou} in the diagonal $K$-matrix can be interpreted
as an algebraic freedom in defining the boundary interaction in the corresponding conformal
boundary loop model \cite{confbound}.

\subsection{Two-colour structure and conserved magnetisations of the $a_3^{(2)}$ model}
\label{section:twocolourstructure}

We now wish to go the opposite way, in order to make transparent the two-colour structure hidden in the vertex formulation of the $a_3^{(2)}$ model. 
First relabel the basis states $\alpha=1,2,3,4$ as $-2,-1,1,2$ (so in particular $\alpha' = -\alpha$), and give these states the following interpretation as the product of $U_q\left(sl_2\right)$ spin-${1 \over2}$ states 
\begin{eqnarray}
 -2,-1,1,2 &=& \left|-\right\rangle_1 \otimes  \left|-\right\rangle_2 , \quad  \left|-\right\rangle_1 \otimes  \left|+\right\rangle_2 , \quad  \left|+\right\rangle_1 \otimes  \left|-\right\rangle_2 , \quad  \left|+\right\rangle_1 \otimes  \left|+\right\rangle_2  \\ 
 \alpha &=& 3 a_1 + a_2 \,,
 \label{aa1a2}
\end{eqnarray}
where $a_1 = S_1^{(z)}$ and $a_2 = S_2^{(z)}$ take values $\pm \frac12$ and will
be interpreted as the $z$-component of spin in each of the two models.
In this formulation the charge $t_{-2,-1,1,2}=-1,0,0,1$ defined earlier can just be interpreted as the total spin, $t_\alpha = S^{(z)}_1 + S^{(z)}_2$. 

We can define on each pair of sites Temperley-Lieb generators in a standard way 
\begin{eqnarray}
 \left(e_1\right)_{a_1 a_2,b_1 b_2}^{c_1 c_2,d_1 d_2} &=& \delta_{a_1+b_1,0}  \delta_{c_1+d_1,0} q^{c_1-b_1} \\
  \left(e_2\right)_{a_1 a_2,b_1 b_2}^{c_1 c_2,d_1 d_2} &=& \delta_{a_2+b_2,0}  \delta_{c_2+d_2,0} q^{c_2-b_2}
\end{eqnarray}
and check that these generators obey the same algebraic relations as those in (\ref{Re1e2}), namely (\ref{TLrelations}).
Let us represent these generators graphically in the loop language of (\ref{TLonecolour}), which allows to write (\ref{RGM}) acting on two sites $a,b$ as
\begin{eqnarray}
\RR_{a,b} &=& P_a P_{b} \left[ w_I \raisebox{-0.4cm}{\Rii} + w_X \left( \raisebox{-0.4cm}{\Rei} + \raisebox{-0.4cm}{\Rif} \right)  + w_E \raisebox{-0.4cm}{\Ref} \right] P_a^{-1} P_{b}^{-1} \\
&\equiv& P_a P_{b} \RR_{\text{loop}} P_a^{-1} P_{b}^{-1}  \,,
\end{eqnarray}
where $w_I$, $w_X$, $w_E$, are coefficients that depend on $\gamma$ and $\lambda$. The $P_{i}$ are gauge factors which amount to multiplying the states $\alpha=\pm 1$ (resp.\ $\alpha=\pm 2$) by $\mathrm{i}$ on odd (resp.\ even) sites, namely
\begin{eqnarray}
P_a &=& \mathrm{diag}\left( 1,\mathrm{i} ,\mathrm{i},1 \right) \otimes \mathbf{1} \equiv U \otimes \mathbf{1} \,, \\
P_b &=& \mathbf{1}  \otimes \mathrm{diag}\left( \mathrm{i},1,1 ,\mathrm{i} \right)  \equiv \mathbf{1}\otimes V \,.
\end{eqnarray} 
Nothing changes conversely if one decides to instead multiplying $\alpha=\pm 1$ (resp.\  $\alpha=\pm 2$) by $\mathrm{i}$ on even (resp.\ odd) sites, which amounts to exchanging $U$ and $V$.

Therefore, considering the $\RR_{\mathcal{A}i}$ matrix acting on the auxiliary space ${\cal A}$ and the quantum space labelled $i$ we can use the equivalence just mentioned to write  
\begin{eqnarray}
 \RR_{\mathcal{A},i} &=& U_i V_{\mathcal{A}} \RR_{\text{loop}} V_i^{-1} U_{\mathcal{A}}^{-1} \quad \mbox{for $i$ even} \,, \\
 \RR_{\mathcal{A},i} &=& V_i U_{\mathcal{A}} \RR_{\text{loop}} U_i^{-1} V_{\mathcal{A}}^{-1} \quad \mbox{for $i$ odd}\,.
\end{eqnarray}
Since the square lattice is bipartite, the factors of $U^{\pm 1}$ and $V^{\pm 1}$ coming from adjacent sites will cancel out when forming the transfer matrix $T_L(\lambda)$, so it is equivalent to express the latter in the form (\ref{T6GM}) with $\check{R}_{\mathcal{A},i} = \check{R}_{\rm loop}$, i.e., with the gauge matrices being omitted. This observation completes the equivalence between the $a_3^{(2)}$ vertex model and the two-colour loop model, up to boundary effects and other subtleties to be discussed in section~\ref{sec:bcs} below.

Just like in the well-known construction in the one-colour (Potts) case (see e.g.~\cite{Richard}), the loop transfer matrix has a block-triangular structure in terms of the number of through-lines (or ``watermelon legs'') $l_1$ and $l_2$ propagating in each of the Potts models. As far as the eigenvalue problem is concerned, it is therefore equivalent to impose the strict conservation of the quantum numbers $(l_1,l_2)$. On the other hand, the vertex-model transfer matrix commutes with both of the total magnetisations, 
$S_1^{(z)}\equiv \sum_{i=1}^L \left(S_1^{(z)}\right)_i$ and
$S_2^{(z)}\equiv \sum_{i=1}^L \left(S_2^{(z)}\right)_i$, and so it can be diagonalised
in sectors of fixed total magnetisation. It follows that the sector of the loop-model
transfer matrix with a fixed number $(l_1,l_2)$ of through-lines of each colour is related with that of the vertex-model transfer matrix with magnetizations $S_1^{(z)}={l_1 \over 2}$ and $S_2^{(z)}={l_2 \over 2}$.

\subsection{The periodic loop model and its associated twisted vertex model}
\label{sec:bcs}

To identify the models completely, that is, for instance, to reformulate the periodic loop transfer matrix in terms of the transfer matrix (\ref{T6GM}), there are however still two aspects that need to be taken care of.

\subsubsection{Choice of the boundary conditions}
\label{section:choiceoftwists}

In the periodic loop model as considered in \cite{FJ08}, there can exist non contractible loops, that is, closed loops that wind horizontally around the periodic direction.
These must have the same weight $n=q+q^{-1}=2\cos\gamma$ as the contractible ones, a fact which needs to be taken in account in the vertex model by choosing correctly the twist in (\ref{T6GM}). Let us write the latter in terms of two independent twist angles $\phi_1$ and $\phi_2$, associated with each of the two colours,
\begin{equation}
 \tau_{\mathcal{A}} = \mathrm{e}^{-2\mathrm{i} \left( \phi_1 s_1^{(z)}+\phi_2 s_2^{(z)}\right)} \,.
 \label{twistmatrix}
\end{equation}
We stress that $s_i^{(z)} = \pm \frac12$ denotes here the local magnetisation along the auxiliary space; it should not be confused with the global magnetisation $S_i^{(z)} = -\frac{L}{2},\ldots,\frac{L}{2}$ on the quantum spaces, a quantity which is conserved by the transfer matrix. 
The proper values to give to the twist angles follows depend on $S_i^{(z)}$. For each of $i=1,2$ we must choose them as follows:
\begin{itemize}
 \item When $S_i^{(z)}=0$ there can exist non contractible loops of colour $i$. The correct choice is $\phi_i=\gamma$, so that each non contractible loop gets a weight $\mathrm{e}^{2\mathrm{i}\gamma{1\over 2}}+\mathrm{e}^{-2\mathrm{i}\gamma{1\over 2}}=n$.  
 \item When $S_i^{(z)} \neq 0$ the presence of through-lines forbids the presence of non contractible loops of colour $i$. The correct choice is then $\phi_i = 0$, since otherwise the through-lines would pick up spurious phase factors when spiraling around the horizontal, periodic direction.
\end{itemize}

\subsubsection{The twisted vertex model as an enlarged periodic loop model}
\label{sec:enlarged}

Even with the correct choice of twist angles, there is a subtle difference between
the twisted vertex model and the periodic loop model. The reason for this is that
the vertex model has a larger space of states.

To show this, we first focus on a single loop colour. Consider as an example
the system of size $L=4$. In the loop model there are $2$ possible states without through-lines which can be represented as graphically as
$\begin{tikzpicture}[scale=0.5]
 \draw[red,line width=1.0] (0,0) arc(180:360:3mm and 5mm);
 \draw[red,line width=1.0] (1.2,0) arc(180:360:3mm and 5mm);
\end{tikzpicture}$
and
$\begin{tikzpicture}[scale=0.5]
 \draw[red,line width=1.0] (0,0) arc(180:360:9mm and 5mm);
 \draw[red,line width=1.0] (0.6,0) arc(180:360:3mm and 2.5mm);
\end{tikzpicture}$.
In the sector $S^{(z)} = 0$ of the vertex model there are obviously ${4 \choose 2} = 6$
states. The difference is that the loop model gives the same weight $n$ to any
loop, contractible or not, whereas the vertex model can control the weight of the
non contractible loop independently by means of the twist angle. To endow the
loop model with the capability of distinguishing between contractible and non
contractible loops, we must enlarge its state space with another 4 states. We
can represent those graphically as
\raisebox{-0.5mm}{$\begin{tikzpicture}[scale=0.5]
 \draw[red,line width=1.0] (0,0) arc(180:360:3mm and 5mm);
 \draw[red,line width=1.0] (1.2,0) arc(180:360:3mm and 5mm);
 \draw[black,fill] (0.3,-0.5) circle(0.5ex);
\end{tikzpicture}$},
\raisebox{-0.5mm}{$\begin{tikzpicture}[scale=0.5]
 \draw[red,line width=1.0] (0,0) arc(180:360:3mm and 5mm);
 \draw[red,line width=1.0] (1.2,0) arc(180:360:3mm and 5mm);
 \draw[black,fill] (1.5,-0.5) circle(0.5ex);
\end{tikzpicture}$},
\raisebox{-0.5mm}{$\begin{tikzpicture}[scale=0.5]
 \draw[red,line width=1.0] (0,0) arc(180:360:9mm and 5mm);
 \draw[red,line width=1.0] (0.6,0) arc(180:360:3mm and 2.5mm);
 \draw[black,fill] (0.9,-0.5) circle(0.5ex);
\end{tikzpicture}$}, and
\raisebox{-0.5mm}{$\begin{tikzpicture}[scale=0.5]
 \draw[red,line width=1.0] (0,0) arc(180:360:9mm and 5mm);
 \draw[red,line width=1.0] (0.6,0) arc(180:360:3mm and 2.5mm);
 \draw[black,fill] (0.9,-0.25) circle(0.5ex);
 \draw[black,fill] (0.9,-0.5) circle(0.5ex);
\end{tikzpicture}$},
where a mark on an arc now means that it has traversed the periodic
boundary condition. The number of marks add up modulo 2 upon
multiple traversals and upon concatenating two arcs through the action
of TL generators.

We shall refer to the loop model where arcs in the sector without through-lines can be marked as the {\em enlarged loop model}. (We do not mark arcs in sectors with through-lines, since there cannot be any non contractible loops anyway.)
The original, periodic loop model, will be in contrast refered to as the {\em original loop model}. We now claim that the enlarged loop model is equivalent to the twisted vertex model, in the sense that their state spaces are isomorphic.

In fact, it is not difficult to establish a bijection between the state spaces. Reading the states of the enlarged loop model from left to right, replace each opening of an unmarked (resp.\ a marked) loop by an up-spin (resp.\ a down-spin) and each closing
by a down-spin (resp.\ an up-spin). In this way, the first two states given above
become
$\uparrow \downarrow \uparrow \downarrow$ and
$\uparrow \uparrow \downarrow \downarrow$, while the latter four states become
$\downarrow \uparrow \uparrow \downarrow$,
$\uparrow \downarrow \downarrow \uparrow$,
$\downarrow \uparrow \downarrow \uparrow$, and
$\downarrow \downarrow \uparrow \uparrow$.
The mapping extends to sectors with through-lines, provided we replace each
through-line by an up-spin. To establish the reverse mapping, consider any given
initial spin. Compute the accumulated magnetisation upon moving rightwards (crossing
the periodic boundary condition if necessary) until the magnetisation becomes zero,
or the same spin is reached again. In the former case, the spin where the magnetisation
becomes zero is linked by an arc to the initial spin. The corresponding spins are obviously opposite, and if the down-spin is to the left of the up-spin (and only if we are in the $S^{(z)}=0$ sector) the arc is marked.
In the latter case, there is no corresponding spin and the initial spin is a through-line.

It is an elementary exercise to show that in the original loop model the sector without through-lines has dimension
\begin{equation}
 d_0(L) = {L \choose L/2} - {L \choose L/2 + 1}
 = \frac{1}{L/2 + 1} {L \choose L/2} \,.
\end{equation}
In the enlarged loop model the sector with $2l$ through-lines has dimension 
\begin{equation}
  d'_{2l}(L) = {L \choose L/2 + l} \,,
\end{equation}
which is obvious because of the equivalence with the vertex model. The
original loop model has the same dimensions in the sectors with through-lines,
i.e., $d_{2l}(L) = d'_{2l}(L)$ for $l \neq 0$.

The extension of these considerations to the two-colour loop model is
obvious, since the two loop colours behave independently. In particular
the total dimension in the sector with $(l_1,l_2)$ through-lines is the product of the dimensions for each of the colours:
\begin{equation}
 d_{(2l_1,2l_2)}(L) = d_{2l_1}(L)  \, d_{2l_2}(L) \,.
\end{equation}

\section{Conformal spectrum of the $a_3^{(2)}$ model: Bethe Ansatz results}

\label{section:BAEa32}

Having cleared up the relationship between the two-colour dense loop model of \cite{FJ08} and the $a_3^{(2)}$ twisted vertex model, we now turn to the Bethe Ansatz study of the latter. The numerical study of the Bethe Ansatz equations allows us in particular to attain the eigenvalues in large finite size, and the close relationship with the $a_2^{(2)}$ model studied in \cite{VJS:a22} will then permit us to infer the conformal spectrum in the continuum limit. 

Our Bethe Ansatz study of the $a_3^{(2)}$ model is part of a broader study of the $a_n^{(2)}$ models, which we plan to develop in a future publication \cite{VJS:an2}.
Each of these models comprises three regimes, denoted I, II and III, as already explained in the $a_2^{(2)}$ case in \cite{VJS:a22}. In the $a_3^{(2)}$ case, and referring to the isotropic models $\lambda = \lambda_\pm$ given by (\ref{lambdaisoRIII})-(\ref{lambdaisoRI}), these three regimes correspond to the following choice of the parameters:
\begin{itemize}
 \item Regime I corresponds to the isotropic point $\lambda_-$ with $\gamma \in \left[0 , {\pi \over 2}\right]$.
 \item Regime II corresponds to the isotropic point $\lambda_+$ with $\gamma \in \left[{\pi \over 4} , {\pi \over 2}\right]$.
 \item Regime III corresponds to the isotropic point $\lambda_+$ with $\gamma \in \left[0 , {\pi \over 4}\right]$.
 \end{itemize}
Only the regime III corresponds to the range of parameters describing the critical point (\ref{RJF}) for $2 < Q \le 4$ and we will therefore not discuss the regimes I and II any longer.

\subsection{Evidence for a non compact boson}

Before entering the details of our numerical results, let us present the main lines of what has convinced us
about the presence of a non compact boson in the continuum limit of regime III. The first piece of 
evidence can be related to the two following observations: 
\begin{itemize}
 \item In the periodic (untwisted) case, the conformal exponents associated with each level show a very slow convergence with the size $L$ of the system.
 \item Turning on the twists $\phi_1$ and $\phi_2$, we observe changes of regimes for these exponents, beyond which these are described by different analytical formulae and the convergence issues observed in the small-twist regime have disappeared.   
\end{itemize}

This is very reminiscent of the features observed in the regime III of the $a_2^{(2)}$ model \cite{Nienhuis},
which lead us in \cite{VJS:a22} to associate the continuum limit of this model with Witten's Euclidean black hole
CFT \cite{BlackHoleCFT,Troost,RibSch} (which can also be considered as the coset $SL(2,\mathbb{R})_k/U(1)$).
It is therefore very tempting to interpret the continuum limit of the $a_3^{(2)}$ model in regime III as another non
compact CFT. A systematic study of this CFT will be postponed to a subsequent publication on the $a_n^{(2)}$
models for general $n$ \cite{VJS:an2}. Instead, we will here rely on the fact that the features described above,
which are highly unusual within the context of ordinary CFTs (for instance those described by a Coulomb gas for
bosons of compact radius), have a natural interpretation within the context of non compact, cigar-like CFTs.

To this end, it is useful to recall some basic features of the black hole CFT \cite{BlackHoleCFT,Troost,RibSch}.
It is written in terms of an action of
two fluctuating fields $r$ and $\theta$, on some target space with metric 
\begin{equation}
 ds^2={k\over 2} d\sigma^2,~d\sigma^2=(dr)^2+\tanh^2r (d\theta)^2 \,.
\end{equation}
This target space is associated  to a two-dimensional surface in three dimensions with the rough shape of a cigar, hence the familiar name `cigar CFT'. More precisely, the target has rotational invariance around the $z$ axis, while the radius in the $x,y$ plane is given by $\tanh r$, where $r\geq 0$ denotes the geodesic distance from the origin.
 The best way to understand the physics of this CFT is to study it within the minisuperspace approximation, that is, solve the Laplacian on the target \cite{RibSch}
\begin{equation}
 \Delta=-{2\over k}\left[\partial_r^2+\left(\coth r+ \tanh r \right)\partial_r+\hbox{coth}^2 r\partial_\theta^2\right] \label{LapTar}\,.
\end{equation}
In this limit, there are no $L^2$-normalisable eigenfunctions. The whole spectrum is obtained from
$\delta$-function normalisable eigenfunctions, which depend on two parameters: one is $n\in \mathbb{Z}$,
the angular momentum of rotations around the axis, and the other, $J=-{1\over 2}+is$, is related to
the momentum $s \in \mathbb{R}$ along the $\rho$-direction of the cigar. 
Each eigenfunction of the Laplacian lifts into a primary state in the CFT, and the corresponding Laplacian eigenvalues read
\begin{equation}
 x=h+\bar{h}=-{2J(J+1)\over k}+{n^2\over 2k} \,.
\end{equation}
The relation between $J$ and $s$ is imposed by the normalisability.

In finite size, the existence of a continuum of primary fields corresponding to various values of $s$ 
is associated to towers of excited transfer matrix eigenstates indexed by an integer $j$ and with
conformal weights of the form 
\begin{equation}
 \Delta_j(L) \sim \mbox{compact part} + j^2 {\frac{A}{[B+\log L]^2}} \,.
 \label{logarithmicscaling}
\end{equation}
There is thus a lattice regularisation of the momentum $s$, which also explains the slow convergence of the corresponding exponents. 

Taking account of the ``stringy'' corrections requires considering non zero winding modes (indexed by the winding number $w$) of strings around the longitudinal direction of the cigar, which could be implemented in the $a_2^{(2)}$ lattice model by varying the twist parameter $\varphi$. A consequence of these corrections was shown in \cite{BlackHoleCFT, RibSch,Troost} to be that, on top of the continuum of normalisable states discussed
above, the theory also admits {\sl discrete states} which can be observed as an additional discrete set of conformal exponents popping out of the continuum beyond some particular value of the twist parameter, hence explaining the changes of regimes and the convergence improvements mentioned above.  

Although the link between non compactness and discrete states was only worked out in details for
the $SL(2,\mathbb{R})_k/U(1)$ case, we wish to sketch here that it has a quite general origin related to the
geometry of the target space, and hence could very probably generalise to other non compact CFTs.
This reproduces in a very evocative way the arguments of \cite{BlackHoleCFT}. The primary fields are labeled by three integers, namely the momentum $n$ and winding number $w$ around the compact direction of the cigar, as well as the momentum $J$ (or $s$) in the non compact direction. Although $n$ and $w$ are treated on the
same footing by the CFT, they lead to very different sigma model descriptions, as for instance non zero values of $w$ are not accounted for in the minisuperspace approach. To described non zero winding modes, one has to perform a duality transformation $(r,\theta) \to (r,\tilde{\theta})$ on the cigar target space, which is turned into  a singular, trumpet-like geometry with metrics
\begin{equation}
 ds^2=(dr)^2+4\coth^2{r \over 2} (d\tilde{\theta})^2 \,.
\end{equation}
This singular new geometry allows for bound states, which are precisely the discrete states referred to above.

In conclusion, the non compactness is closely linked to the existence of discrete states. The non compactness
itself leads directly to the logarithmic scaling (\ref{logarithmicscaling}), which is hard to extract quantitatively
from finite-size numerical data, beyond the observation that the scaling dimensions converge very slowly.
However, the emergence of well-converged discrete states beyond certain values of the twist is quite easy
to detect numerically. Our affirmation that the $a_3^{(2)}$ model contains non compact features is therefore
based on the combined observation of slow convergence compatible with (\ref{logarithmicscaling}) for small
twists, and the emergence of well-converged
discrete states.

\subsection{The twisted Bethe Ansatz equations}

The Bethe ansatz equations for the purely periodic (untwisted) model --- i.e., with $\tau_{\mathcal{A}}= {\rm Id}_{4\times 4}$ in (\ref{T6GM}) --- were derived in \cite{GalleasMartins04}. They involve two different types of roots, which we note $\lambda_i$ ($i=1,\ldots,m_1$) and $\mu_i$ ($i=1,\ldots,m_2$). 

In appendix \ref{app:twistedBAE}, we revisit this derivation with two goals in mind : 
\begin{itemize}
\item First, we wish to understand precisely how the numbers $m_1$ and $m_2$ of each kind of roots are related to the different sectors of fixed magnetisation in a system of size $L$
\item Second, we need to slightly extend the working of \cite{GalleasMartins04}, since we are interested not only in the periodic case, but also in its generalisation to an arbitrary twist angles $\phi_1$ and $\phi_2$. 
\end{itemize}

As a result of this analysis, the numbers $m_1$ and $m_2$ are seen to be related to the magnetizations $S_1^{(z)}$ and $S_2^{(z)}$ by 
\begin{eqnarray}
 S_1^{(z)} &=& -{L \over 2} +  m_1 - m_2  \,, \nonumber \\
 S_2^{(z)} &=& -{L \over 2} + m_2 \,.
 \label{main:relateSandm}
\end{eqnarray}

The twisted Bethe equations read \cite{MartinsNienhuis}

\begin{eqnarray}
\mathrm{e}^{2\mathrm{i}\phi_1}\left(\frac{\sinh (\lambda_i-\mathrm{i}\frac{\gamma}{2})}{\sinh (\lambda_i+\mathrm{i}\frac{\gamma}{2})} \right)^L &=& 
\prod_{j=1, j\neq i}^{m_1} \frac{\sinh (\lambda_i-\lambda_j-\mathrm{i}\gamma)}{\sinh (\lambda_i-\lambda_j+\mathrm{i}\gamma)} 
\prod_{k=1}^{m_2} \frac{\sinh (2(\lambda_i-\mu_k+\mathrm{i}\frac{\gamma}{2}))}{\sinh (2(\lambda_i-\mu_k-\mathrm{i}\frac{\gamma}{2}))}  \nonumber \\
\mathrm{e}^{2\mathrm{i}\left(\phi_1 - \phi_2\right)} \prod_{j=1}^{m_1} \frac{\sinh (2(\mu_k-\lambda_j-\mathrm{i}\frac{\gamma}{2}))}{\sinh (2(\mu_k-\lambda_j+\mathrm{i}\frac{\gamma}{2}))} &=& 
\prod_{l=1, l\neq k}^{m_2} \frac{\sinh (2(\mu_k-\mu_l-\mathrm{i}\gamma))}{\sinh (2(\mu_k-\mu_l+\mathrm{i}\gamma))} \,,
\end{eqnarray}
and the corresponding eigenvalues, in the notations of \cite{GalleasMartins04}, are 
\begin{eqnarray}
 \Lambda^{(4)}(\lambda) &=& \mathrm{e}^{\mathrm{i} \left( \phi_1 + \phi_2 \right)} \left[a_{1}(\lambda)\right]^L \frac{Q_1\left( \lambda + \mathrm{i}\frac{\gamma}{2} \right)}{Q_1\left( \lambda - \mathrm{i}\frac{\gamma}{2} \right)}
 +  \mathrm{e}^{\mathrm{i} \left( -\phi_1 - \phi_2 \right)} \left[d_{4,4}(\lambda)\right]^L \frac{Q_1\left( \lambda - \mathrm{i}\frac{5\gamma}{2} + \mathrm{i}\frac{\pi}{2} \right)}{Q_1\left( \lambda - \mathrm{i}\frac{3\gamma}{2} + \mathrm{i}\frac{\pi}{2} \right)}  \nonumber \\
 & & +  \left[b(\lambda)\right]^L \left( \mathrm{e}^{\mathrm{i} \left( \phi_1 - \phi_2 \right)} G_1\left(\lambda  \right)  + \mathrm{e}^{\mathrm{i} \left(- \phi_1 + \phi_2 \right)} G_2\left(\lambda  \right) \right) \,,
\end{eqnarray}
where $Q_1(\lambda) \equiv \prod_{i=1}^{m_1} \sinh\left(\lambda - \lambda_i\right)$ and $Q_2(\lambda) \equiv \prod_{i=1}^{m_2} \sinh\left(\lambda - \mu_i\right)$.

As usual in integrable systems, the Hamiltonian of the corresponding 1D chain can be obtained from the transfer matrix by taking the very anisotropic limit, 
\begin{equation} 
 H^{(L)} =  \mp\left.\frac{\mathrm{d}}{\mathrm{d}\lambda}\log T^{(L)}(\lambda) \right|_{\lambda=0} \,,
\end{equation}
which allows to rewrite its eigenvalues in terms of the Bethe roots 
\begin{equation}
 E =  \pm\sum_{i=1}^{m_1} \frac{2 \sin \gamma}{2\cosh 2\lambda_i - \cos\gamma} \,.
 \label{Energy}
\end{equation}
The two possible signs for the eigenenergies $E$ of the quantum hamiltonian $H^{(L)}$ produce two different regimes for the low-lying excitations. 
In terms of the transfer matrix eigenvalues, the plus sign in (\ref{Energy}) has its low-lying spectrum corresponding to the leading eigenvalues at the isotropic point (\ref{lambdaisoRIII}) while the minus sign corresponds to the leading eigenvalues at (\ref{lambdaisoRI}). This explains the subscripts of $\lambda_\pm$ used in (\ref{lambdaisoRIII})--(\ref{lambdaisoRI}).

\subsection{Low-lying spectrum at zero twist in regime III}

We first consider the conformal spectrum in the untwisted case, before turning on the twist in
the following section.

\subsubsection{Classification of the low-lying excitations}

\begin{figure}
\begin{center}
 \includegraphics[width=90mm,height=70mm]{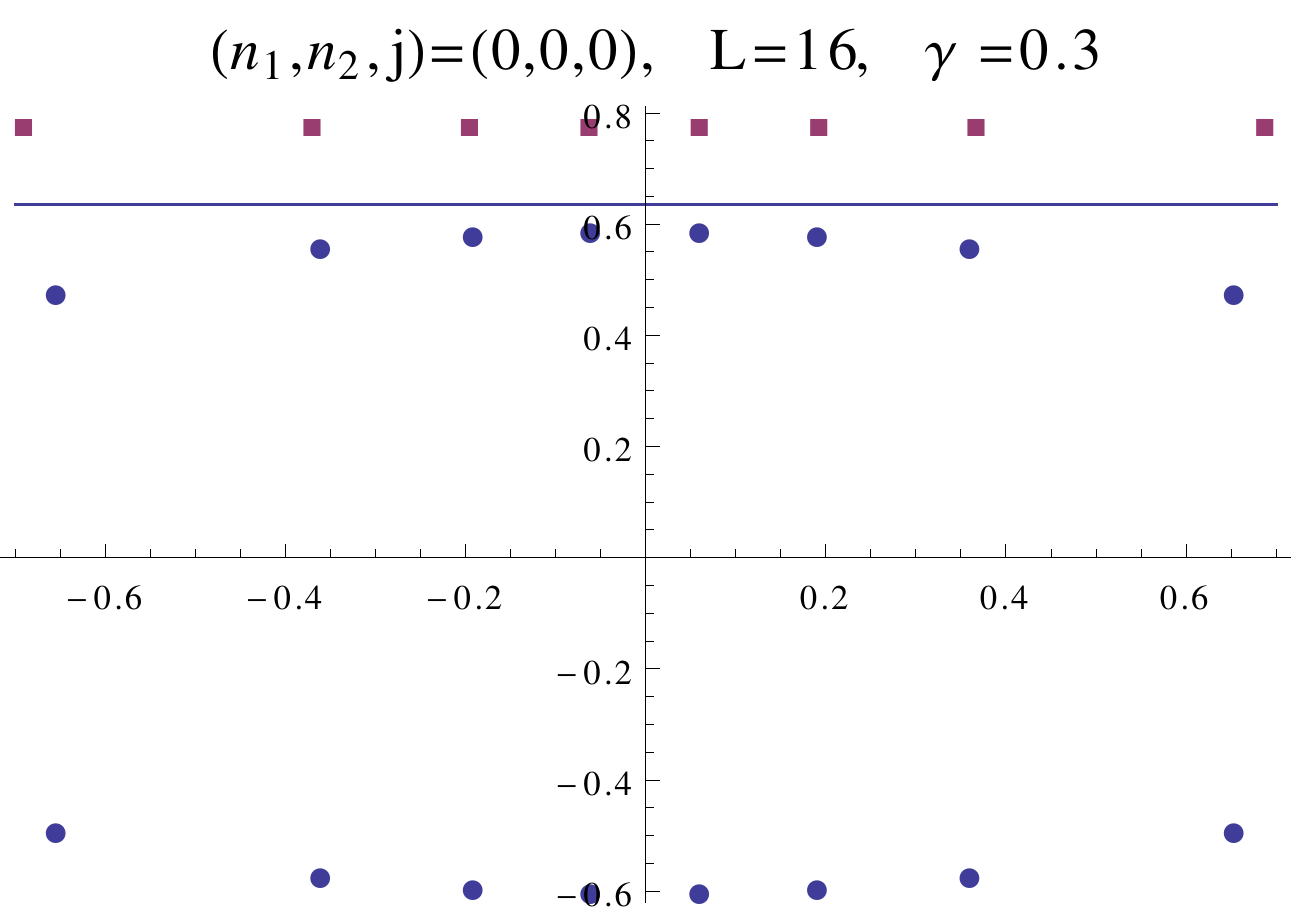}
 \end{center}
\caption{Configuration of the $\lambda$ (in blue) and $\mu$ (in purple) roots corresponding to the ground state of regime III in the $n_1 = n_2=0$ sector, at $\gamma = \frac{3}{10}$ and for a system size $L=16$. We also plotted the line of imaginary part $\left({\pi \over 4} - {\gamma \over 2}\right)$, for comparison.}                                 
\label{fig:E000g030L16}                          
\end{figure}

The regime III, in which lies the critical point (\ref{RJF}) for $k>2$, corresponds to the plus sign in the definition of the energy (\ref{Energy}) and the isotropic point (\ref{lambdaisoRIII}), and to $\gamma \in \left[ 0, {\pi \over 4} \right]$.
In this regime we found (numerically, by comparison with exact diagonalization of the transfer matrix and use of the Mc Coy method \cite{VJS:a22} for sizes up to $L=12$) that the ground state is described by a sea of $L \over 2$ 2-strings (pairs of conjugate roots) of $\lambda$ roots, with imaginary parts close to $\pm \left({\pi \over 4} - {\gamma \over 2}\right)$, together with a sea of $\mu$ roots with imaginary part precisely $\pi \over 4$. These are represented for $L=16$ in figure \ref{fig:E000g030L16}.

We now describe the classification of the low-lying excitations with respect to this ground state.

First, we point out that in the untwisted, periodic case, the transfer matrix commutes with the momentum operator $P^{(L)} = T^{(L)}(0)$, which acts on the states as a unit translation. The eigenstates can therefore be classified according to their momentum eigenvalue, defined modulo $L$. As usual in such systems, the ground state and lowest-lying levels in each sector of given magnetisation have zero momentum, i.e., they are translationally invariant. 
We will restrict to such states in this discussion, and refer to our general work on $a_n^{(2)}$ \cite{VJS:an2} for the conformal weights associated with states of non zero momenta. 

The zero momentum excitations can be labeled by three integers, $(n_1, n_2, j)$. The first two correspond respectively to the magnetizations $S_{1}^{(z)}$ and $S_2^{(z)}$, and the last one is an extra index labeling the level of different excitations in a given magnetisation sector, in a sense that we shall make precise now. 

\begin{paragraph}{Excitations in the $n_1 = n_2 = 0$ sector.} 
All these excitations have $m_1 = L$ and $m_2 = {L \over 2}$.
The excitation $(0,0,j)$, $j=1,2,\ldots$ is obtained from the ground state ($j=0$) roots configuration by replacing $j$ 2-strings of $\lambda$-roots by the same number of antistrings, that is, of pairs of anticonjugate ($\equiv$ having opposite real parts) roots with imaginary part $\pi \over 2$.
\end{paragraph}

\begin{paragraph}{Ground states and excitations in the other sectors.} 
For general $(n_1 , n_2)$ we now have $m_1 = L - n_1- n_2$ and $m_2 = {L \over 2}-n_2$, by (\ref{main:relateSandm}) and after an immaterial sign change of the magnetisations.

More precisely, the roots configurations corresponding to the ground states in these sectors involve
$m_2 = {L \over 2}-n_2$ $\mu$-roots with imaginary part ${\pi \over 4}$, the same number of
2-strings for the $\lambda$-roots, whereas the remaining $\lambda$ roots align on the axis of imaginary part $\pi \over 2$ (see appendix \ref{app:RootsConfigs}).

Just as in the $(n_1,n_2)=(0,0)$ sector, the $j$th excited state is obtained
by replacing $j$ 2-strings by antistrings of imaginary part $\pi \over 2$.
\end{paragraph}

\subsubsection{Conformal spectrum of the untwisted chain}

It is well-known from conformal field theory that the scaling of the energies with the size $L$ allows one to extract the conformal spectrum. 
The finite-size scaling of the ground state energy yields the central charge, 
\begin{equation}
 E_{0,0,0}(L) = E_{\infty} - v_{\rm F} \frac{\pi c}{6 L^2} + O\left(1 \over L^4 \right) \,,
\end{equation}
whereas the scaling of the gap between the ground state and the different excited levels yields the corresponding conformal weights $x_{n_1,n_2,j} = \Delta_{n_1,n_2,j} + \bar{\Delta}_{n_1,n_2,j}$ 
via
\begin{equation}
 E_{n_1,n_2,j}(L) -  E_{0,0,0}(L)  =  v_{\rm F} \frac{2 \pi x_{n_1,n_2,j}}{L^2} + O\left(1 \over L^4 \right) \,.
\end{equation}
In these two formulae $v_{\rm F}$ is the Fermi velocity, which can be found from the scattering equations in the continuum limit to be
\begin{equation}
 v_{\rm F} (\gamma) = {\pi \over \pi - 4 \gamma} \,. 
\end{equation}

In the sequel it will turn out convenient to work with the effective central charges, rather than with the conformal weights, associated with each level:
\begin{equation}
 c_{n_1,n_2,j} \equiv c - 12 x_{n_1,n_2,j} \,,
\end{equation}
where we have set $x_{0,0,0}=0$ for the ground state. 

Similarly to what was observed in regime III of the $a_2^{(2)}$ model \cite{VJS:a22}, we find here
\begin{equation}
 -{c_{n_1,n_2,j} \over 12} = x_{n_1,n_2,j} - {c \over 12} = \frac{\gamma}{2\pi}\left( n_1 + n_2 \right)^2 + \left(N_{n_1,n_2,j}\right)^2 \frac{A(\gamma)}{\left[B_{n_1,n_2,j}(\gamma) + \log L \right]^2}  \,,
 \label{eq:ciuntwisted}
\end{equation}
with quite strong numerical support for the following conjectures:
\begin{eqnarray}
 A(\gamma) &=& 10 {\gamma (\pi - \gamma) \over (\pi - 4 \gamma)^2} \,, \\
 N_{n_1,n_2,j} &=& 1 + \left[\mbox{number of $\lambda$-roots with imaginary part $\pi \over 2$}\right] \,. 
 \label{ceffnontwisted}
\end{eqnarray}
The numerical support for the functional dependence of $A(\gamma)$ on $\gamma$ is very strong,
whereas the determination of the proportionality factor $10$ has more moderate support.
The precise determination of the $B_{n_1,n_2,j}(\gamma)$ functions was however beyond the scope of our numerical accuracy, and progress would presumably require solving the non-linear integral equations (NLIE), e.g., along the lines of \cite{CanduIkhlef} for a cognate but simpler model.

The interpretation of (\ref{ceffnontwisted}) is similar to that made in \cite{VJS:a22}: the last term on the right-hand side accounts for a continuous degree of freedom in the continuum limit, or in other terms for a non compact direction in the target space of the corresponding field theory. In other words, the
continuum limit of the $a_3^{(2)}$ model consists of two compact bosons (corresponding to each of the two magnetisations $n_i$, i.e., originating from each of the two Potts models) and one non compact boson (corresponding to the quantum number $j$, i.e., emerging from the non-trivial coupling of the two models).

\subsection{General twist angles}

In order to make the connection with the loop formulation, we now wish to study what happens to the conformal spectrum when the twist angles $\phi_1$ and $\phi_2$ given non-zero values.
We recall that the equivalence between the $a_3^{(2)}$ vertex model and the two-colour loop
model requires $(\phi_1,\phi_2)$ to take the particular sector-dependent values given in
section~\ref{section:choiceoftwists}. Before specialising to that case we shall however study
the conformal weights of the $a_3^{(2)}$ model for general twists.

Turning on a twist involves --- not unexpectedly --- qualitative changes in the roots configuration describing the excitations $(n_1, n_2,j)$ (some of the corresponding roots patterns are described in appendix \ref{app:RootsConfigs}). What is less usual, however, is that we also observe changes of regimes in the central charge and conformal weights. These changes of regimes, which are not crossovers and also do not necessarly coincide with the change of regimes in the roots configurations, are very similar to what we observed in the $a_2^{(2)}$ case \cite{VJS:a22} and can be explained in terms of the so-called 'discrete states' \cite{BlackHoleCFT,Troost,RibSch}.

\subsubsection{Twist and discrete states: a review of the $a_2^{(2)}$ case}

Before detailing our results on the effective central charges, we briefly review a few relevant results  \cite{VJS:a22} on the closely related $a_2^{(2)}$ model.

The $a_2^{(2)}$ model is also defined in terms of a parameter $\gamma$, but regime III corresponds to the range $\gamma \in [0,\frac{\pi}{3}]$ in that case. Its continuum limit is that of one (not two) compact boson and one non compact boson. Accordingly,
the labelling of the low-lying excitations involves only two integers, $n$ and $j$, where $n$ corresponds to the magnetisation. Similarly the twist is defined in terms of just one angle $\phi$. We found in \cite{VJS:a22} that at $\phi=0$ and in the continuum limit the conformal weights related to the states $j=0,1,\ldots$ and fixed $n$ form a continuum, related to a non compact direction of the corresponding sigma model. 

As already observed by Nienhuis {\em et al.} \cite{Nienhuis}, turning on the twist brings along a change of regime for the central charge at $\phi = \gamma$, so that $c$ is described by two different analytical expressions that are tangent at $\phi=\gamma$. The analytical expression for $c$ with $\phi \geq \gamma$ is the largest for all values of $\phi$, but yet the corresponding state is not observed as the ground state in the regime of small twist $\phi \in [0,\gamma]$. The field-theoretical explanation of this fact is that the corresponding state is non normalisable for $\phi \in [0,\gamma]$. We intrepret this physically as a {\em discrete state} that pops out of the continuum (and becomes normalisable) at $\phi = \gamma$.

Similar phenomena hold true for all the states $(n,j)$, resulting in a set of discrete states that one after the other pop out of the continuum when the twist angle $\phi$ passes through appropriate discrete values. The maximum number of discrete states depend on the value of $\gamma$. As $\phi$ approaches $2\pi$ the opposite phenomenon occurs: one after the other the discrete states reintegrate the continuum (and become non normalisable again). These processes are illustrated in Figure~5 of \cite{VJS:a22}.

We shall now see that the same type of processes occur in the $a_3^{(2)}$ case, and will trust our understanding of the $a_{2}^{(2)}$ case to give these processes a similar interpretation. As usual, we defer the field theoretical description to a subsequent publication \cite{VJS:an2}.

\subsubsection{Conformal weights}

Taking several proportionality constants $\alpha$ between $\phi_1$ and $\phi_2$, we increased both twists at fixed $\alpha$, allowing us to conjecture the full $(\phi_1,\phi_2)$-dependence of the central charges $c_{n_1, n_2, j}$. The support for these conjectures is provided both by the numerical solution of the Bethe Ansatz equations themselves and by the formal similarities with the extensively studied $a_2^{(2)}$ case \cite{VJS:a22}.

Note that all eigenvalues, hence all central charges and conformal weights, are even functions of $\phi_1$ and $\phi_2$, allowing us here to consider only $\phi_1, \phi_2 \geq 0$. In the formulae that follow, it is therefore understood that $\phi_1$ and $\phi_2$ are just short-hand notations for $|\phi_1|$ and $|\phi_2|$.

\begin{figure}
\begin{center}
 \includegraphics[width=100mm,height=80mm]{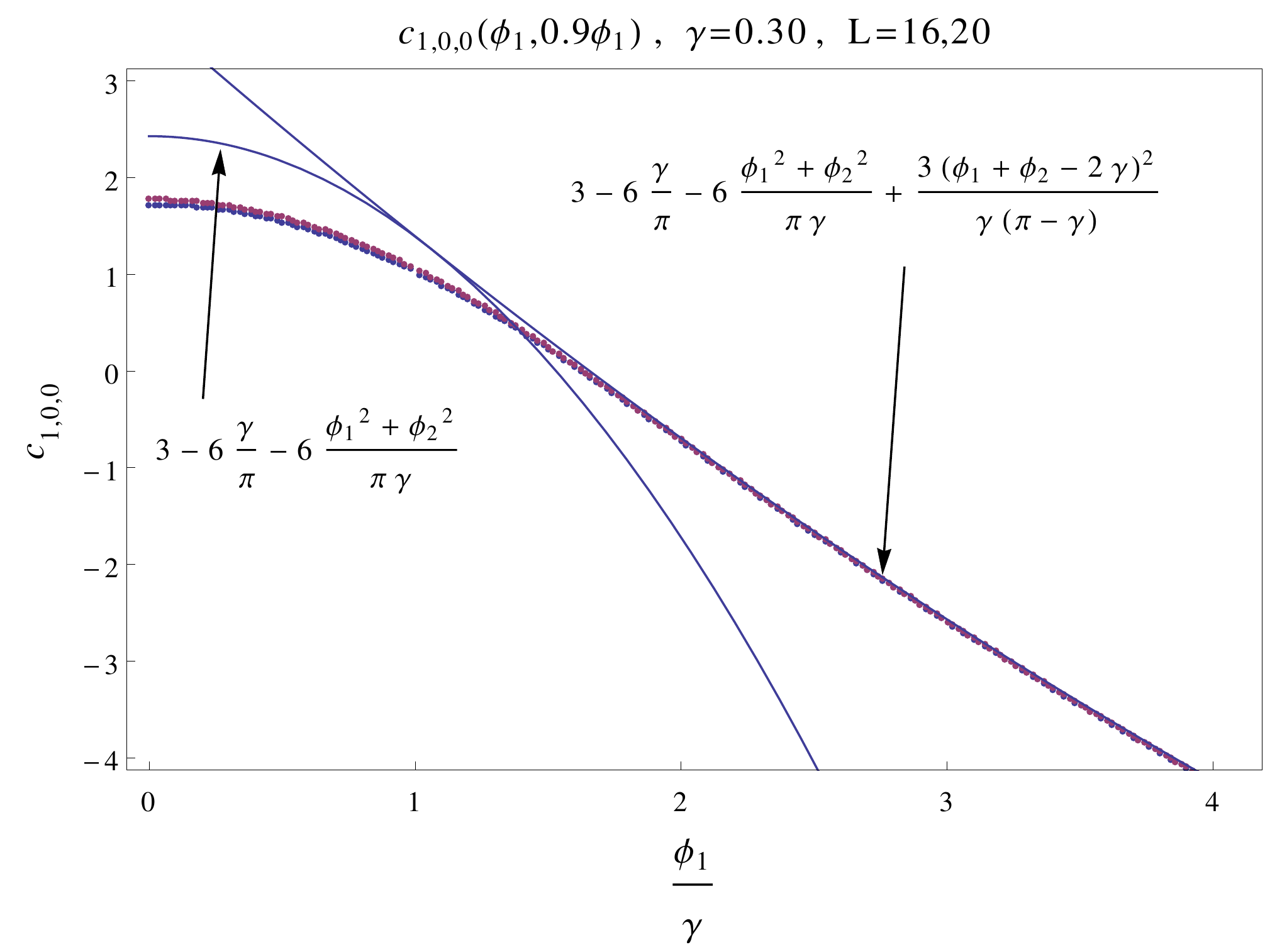}
 \end{center}
\caption{Effective central charge in the sector $(n_1,n_2) = (1,0)$ for $\gamma=\frac{3}{10}$, as a function of $\phi_1$ (with $\phi_2 = \frac{9}{10} \phi_1)$. The solid lines are the conjectured expressions.
}
\label{fig:c100_twist09}                          
\end{figure}

As an example of the numerical accuracy of the results we show in figure \ref{fig:c100_twist09} the central charge $c_{1,0,0}$ measured at $\gamma=\frac{3}{10}$ for $L=16$ and $20$. 
Note in particular that for small twist (here $\phi_1 < \frac{19}{10} \gamma$; see below) the convergence to the first analytical expression --- corresponding to the continuous part of the spectrum --- is really slow, whereas for bigger twist one observes a convergence to the second analytical expression --- corresponding to a discrete state --- that is fast, and similar to what is usually
observed for models with compact continuum limits.

\begin{figure}
\begin{center}
 \includegraphics[width=100mm,height=80mm]{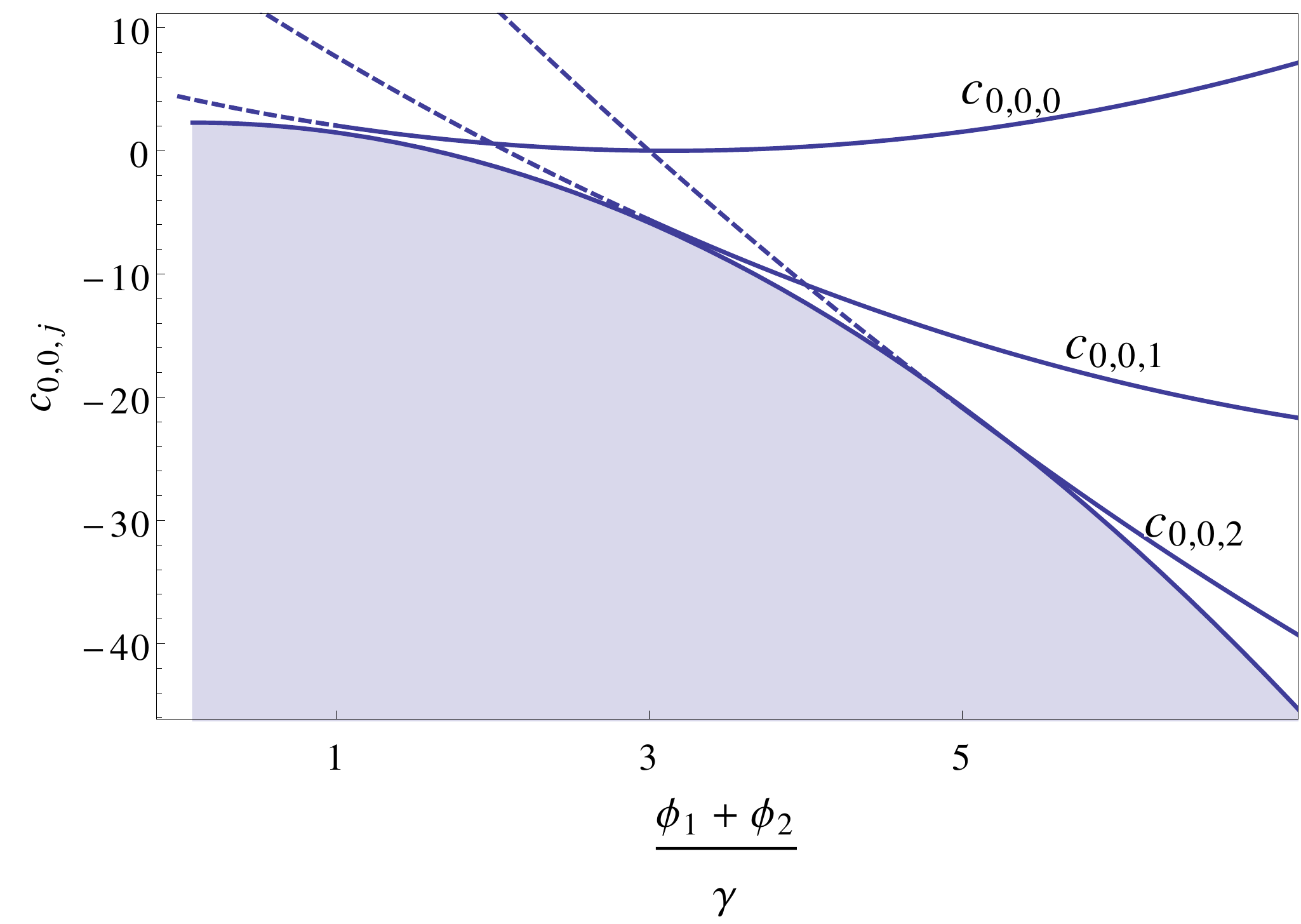}
 \end{center}
\caption{Effective central charges for the first excited states in the sector $n_1 = n_2 = 0$ as a function of the twist angles, for fixed $\gamma$. 
The shaded zone is the continuum, and we represent by dashed curves the analytic continuations of the discrete states central charges $c_{0,0,j}$ to the domain $\phi_1 + \phi_2 \leq (2j+1)\gamma$, where the corresponding states are not normalisable.
}                                 
\label{fig:c00itwisted}                          
\end{figure}

We however first turn to the excited states in the $(n_1,n_2) = (0,0)$ sector, where our numerical 
results lead us to the following conjecture 
\begin{equation}
c_{0,0,j} = \begin{cases}  3 - 6 \frac{\left(\phi_1\right)^2 + \left(\phi_2\right)^2}{\pi \gamma} + o(1) & \mbox{for } \phi_1 + \phi_2 \leq (2 j + 1)\gamma \,,
\\  3 - 6 \frac{\left(\phi_1\right)^2 + \left(\phi_2\right)^2}{\pi \gamma} + 3 \frac{\left(\phi_1 + \phi_2-(2 j + 1)\gamma\right)^2}{\gamma(\pi - \gamma)} & \mbox{for } \phi_1 + \phi_2 \geq (2 j + 1)\gamma \,.
\end{cases}
\label{c00iconj}
\end{equation}
In the first expression we denoted by $o(1)$ the contribution to the central charges of the non-compact degree of freedom, vanishing as $\left( \log L \right)^{-2}$, cf.~(\ref{eq:ciuntwisted}).
We insist on the fact that in the second expression this logarithmic term has disappeared, according to the fact that the corresponding state has lifted from the continuum, and is now a proper discrete state.
This is summed up in figure \ref{fig:c00itwisted}, where we schematically represented the central charges $c_{0,0,j}$ and the corresponding continuum.

In the remainder of this paper we shall use the symbol $o(1)$ with this same meaning, namely
indicating logarithmic corrections due to the non compact boson.

Now going back to the ground states in the different magnetisation sectors, we were led to the following conjectures
\begin{equation}
 c_{n_1,n_2,0} = \begin{cases}  3- 6 {\left( \left(n_1\right)^2 + \left(n_2\right)^2 \right)\gamma \over \pi}  - 6 \frac{\left(\phi_1\right)^2 + \left(\phi_2\right)^2}{\pi \gamma} + o(1) & \mbox{for } \phi_1 + \phi_2 \leq \left(|n_1|+|n_2|+1\right)\gamma \,,
\\  3 - 6 {\left( \left(n_1\right)^2 + \left(n_2\right)^2 \right)\gamma \over \pi} - 6 \frac{\left(\phi_1\right)^2 + \left(\phi_2\right)^2}{\pi \gamma} + 3 \frac{\left(\phi_1 + \phi_2-\left(|n_1|+|n_2|+1\right)\gamma\right)^2}{\gamma(\pi - \gamma)} & \mbox{for } \phi_1 + \phi_2 \geq \left(|n_1|+|n_2|+1\right)\gamma \,.
\end{cases}
\label{cnn0conj}
\end{equation}

We further conjecture that the general formula for $c_{n_1, n_2, j}$, that contains (\ref{c00iconj})--(\ref{cnn0conj}) as special cases, should be
\begin{equation}
 c_{n_1,n_2,j} = \begin{cases}  c^*_{n_1,n_2}  + o(1) & \mbox{for } \phi_1 + \phi_2 \leq \left(|n_1|+|n_2|+2j+1\right)\gamma \,,
\\   c^*_{n_1,n_2} + 3 \frac{\left(\phi_1 + \phi_2-\left(|n_1|+|n_2|+2j+1\right)\gamma\right)^2}{\gamma(\pi - \gamma)} & \mbox{for } \phi_1 + \phi_2 \geq \left(|n_1|+|n_2|+2j+1\right)\gamma \,.
\end{cases}
\label{cnnjconj}
\end{equation}
where we have defined 
\begin{equation}
 c^*_{n_1,n_2} = 3- 6 {\left( \left(n_1\right)^2 + \left(n_2\right)^2 \right)\gamma \over \pi}  - 6 \frac{\left(\phi_1\right)^2 + \left(\phi_2\right)^2}{\pi \gamma} \,.
 \label{cnnstar}
\end{equation}

\section{Conformal spectrum of the dense two-colour loop model}
\label{section:confloop}

By now, we have completed (up to a subtlety that will be explained shortly) our understanding of the $a_{3}^{(2)}$ conformal spectrum in both the twisted and untwisted cases. We are thus ready to come back to the loop model and its exponents.

Magnetic excitations in loop models are described by the so-called watermelon excitations,
corresponding to imposing a fixed number of through-lines. In the context of the two-colour loop model of
interest here, the watermelon operators were defined in section 3.2 of \cite{FJ08}, as operators ${\cal O}_{l_1,l_2}$ that act as sources (or sinks) for a given number $(l_1,l_2)$ of through-lines of each loop colour. The watermelon exponents $x_{l_1,l_2}$ are the critical exponents governing the asymptotic decay of the two point correlation functions,
\begin{equation}
 \langle {\cal O}_{l_1,l_2}({\bf x}_1) {\cal O}_{l_1,l_2}({\bf x}_2) \rangle \sim
 \frac{1}{|{\bf x}_1 - {\bf x}_2|^{2 x_{l_1,l_2}}} \,.
\end{equation}
Although energy-type excitations are also
of interest, we limit ourselves to the study of the watermelon exponents in what follows.

The central charge and first few watermelon exponents of what we refer to as the original loop model (see section~\ref{sec:enlarged}) were measured numerically in \cite{FJ08}.
These authors worked by numerically diagonalising the transfer matrix for sizes up to $L=16$,
in several different sectors $(l_1,l_2)$, but failed to obtain a general formula for the watermelon exponents. Our goal in this final part is to use the results of section \ref{section:BAEa32} to obtain this general expression.
With hindsight, we can state that the difficulties encountered in \cite{FJ08} are 
due to the very particular (logarithmic) finite-size behaviour of the scaling levels entailed by the
non compact boson. Section~\ref{FJcomp} contains an {\em a posteriori} comparison of
our numerical analysis with the one made in \cite{FJ08}.

The watermelon exponent $x_{l_1,l_2}$ associated with the operator ${\cal O}_{l_1,l_2}$ inserting $l_1$ blue lines and $l_2$ red lines can be expressed as
\begin{equation}
 x_{l_1,l_2} = \frac{c - c_{l_1,l_2}}{12} \,,
 \label{linkxtoc}
\end{equation}
where $c_{l_1,l_2}$ is the effective central charge in the loop model's $(l_1, l_2)$-legs sector, and $c$ the central charge. We have seen that the loop model can only be interpreted as a Potts model when $L$ is even. Since $(l_1,l_2)$ must necessarily have the same parity as $L$, we shall henceforth set $l_1 = 2 n_1$ and $l_2 = 2 n_2$. As far as the two-colored loop model is concerned, one could also consider the ``twisted sector'' of odd $L$, and hence odd values of $l_1$ and $l_2$, but we shall refrain from doing so.

To obtain the loop model's central charge and watermelon exponents, the first thing we need to do is therefore to identify the ground states in the different sectors within the bigger spectrum of the extended loop (twisted vertex) model, which is the one we understand from the Bethe Ansatz.
The two models' (isotropic) transfer matrices can be implemented numerically, and their eigenvalues obtained by direct diagonalisation. The general conclusion we draw from the analysis of these eigenvalues are the following:
\begin{itemize}
 \item In sectors with $(0,0)$ legs or with $(l_1\neq 0,l_2 \neq 0)$ legs, the ground states of the two models coincide all through regime III.
 \item In sectors with $(l_1\neq 0,0)$ or $(0,l_2 \neq 0)$ legs, the two ground states do not necessarily coincide. Moreover we observe crossovers. We will make this observation more precise in the following (see section \ref{Section:xl10}).
\end{itemize}

Note that the way in which we have defined the excited states $j \neq 0$ for each sector in the vertex model entails that these also present in the original loop model. In other words, the vertex model
obviously contains more state than the original loop model, but those are not of the form $(n_1,n_2,j)$.
In any case, since the calculation of the central charge and watermelon exponents only involves the ground states in each sector we do not need to consider the excited states any further.

\subsection{Central charge and first set of watermelon exponents}
\label{section:cxnn}

According to the above discussion, we first focus on the sectors with $(0,0)$ legs or with $(l_1\neq 0,l_2 \neq 0)$ legs, that is the sectors $n_1 = n_2 = 0$ and $(n_1\neq 0,n_2 \neq 0)$ in the vertex model, for which the ground states of the original loop model and that of the (appropriately twisted) vertex model coincide. 

We start by considering the central charge $c$, associated with the ground state in the sector $n_1 = n_2 = 0$. As explained in section \ref{section:choiceoftwists}, it is obtained by imposing the twists $\phi_1 = \phi_2 = \gamma = \frac{\pi}{k+2}$ on the vertex model, yielding 
\begin{equation}
 c = c_{0,0,0}(\gamma,\gamma) = 3 -  12{\gamma \over \pi} + 3 {\gamma \over \pi - \gamma} = {3 k^2 \over (k+1)(k+2)} \,,
 \label{eq:measurec}
\end{equation}
which is exactly the expression (\ref{cJF}).

We now turn to the watermelon exponents $x_{l_1\neq 0, l_2 \neq 0}$.
From all that precedes, and in particular from the discussion in section \ref{section:choiceoftwists}, we know that the effective central charge of the loop model, in the sector $l_1=2n_1\neq 0, l_2=2n_2 \neq 0$, is obtained as 
\begin{equation}
 c_{l_1,l_2} = c_{2 n_1,2 n_2} = c_{n_1,n_2,0}(0,0) = 3 - 6 \frac{\left(  \left(n_1\right)^2 + \left(n_2\right)^2   \right)\gamma}{\pi} \,,
\end{equation}
and by (\ref{linkxtoc}) the corresponding watermelon exponent is
\begin{eqnarray}
 x_{2 n_1, 2 n_2} &=& \frac{\left(n_1\right)^2 + \left(n_2\right)^2-2}{2}\frac{\gamma}{\pi} + {\gamma \over  4(\pi - \gamma)} \nonumber \\
 &=& \frac{\left(n_1\right)^2 + \left(n_2\right)^2-2}{2}\frac{1}{k+2} + {1 \over  4(k +1)} \,. 
 \label{watermelon_both_nonzero}
\end{eqnarray}

\subsection{Comparison with the analysis of Ref.~\cite{FJ08}}
\label{FJcomp}

\begin{table}
\begin{center}
\begin{tabular}{l|lll}
 Exponent & $k=3$    & $k=4$    & $k=5$ \\ \hline
  $x_{2,2}$ &0.0625& 0.05& 0.0416667  \\
  $x_{4,2}$ &0.3625& 0.3& 0.255952   \\
 $x_{4,4}$ &0.6625& 0.55& 0.470238   \\
 $x_{6,2}$ & 0.8625& 0.716667& 0.613095 \\
 $x_{6,4}$ &1.1625& 0.966667& 0.827381 \\
 $x_{6,6}$ &1.6625& 1.38333& 1.18452 \\
\end{tabular}
\end{center}
\caption{Numerical values of the watermelon exponent (\ref{watermelon_both_nonzero}).}
\label{tab:watermelon}
\end{table}

In Table \ref{tab:watermelon} we give some numerical values corresponding to (\ref{watermelon_both_nonzero}). This can be directly compared with
Table 1 of \cite{FJ08} that reports the $L \to \infty$ extrapolated results (with indicative error bars)
based on transfer matrix diagonalisations for systems of size $L \le 16$.
The disagreement is in most cases quite spectacular, the discrepancy with the exact results
of our Table~\ref{tab:watermelon} often being more than 50 times the perceived error bar
in \cite{FJ08}.%
\footnote{In particular, our results $x_{2,2} = \frac{1}{4(k+1)}$ and
$x_{4,2} = \frac{8+7k}{4(k+1)(k+2)}$
infirm the conjectures $\frac{4}{(k+1)(k+2)}$ and $\frac{16}{(k+1)(k+2)}$ proposed in \cite{FJ08}.}
The reason for this is obviously the very slow convergence of $c_{n_1,n_2,0}(0,0)$ due to the presence of the logarithmic term at zero twist (\ref{eq:ciuntwisted}), a phemenon that the authors
of \cite{FJ08} had clearly no reason to suspect. 

\begin{figure}
 \begin{center}
  \includegraphics[width=150mm,height=120mm]{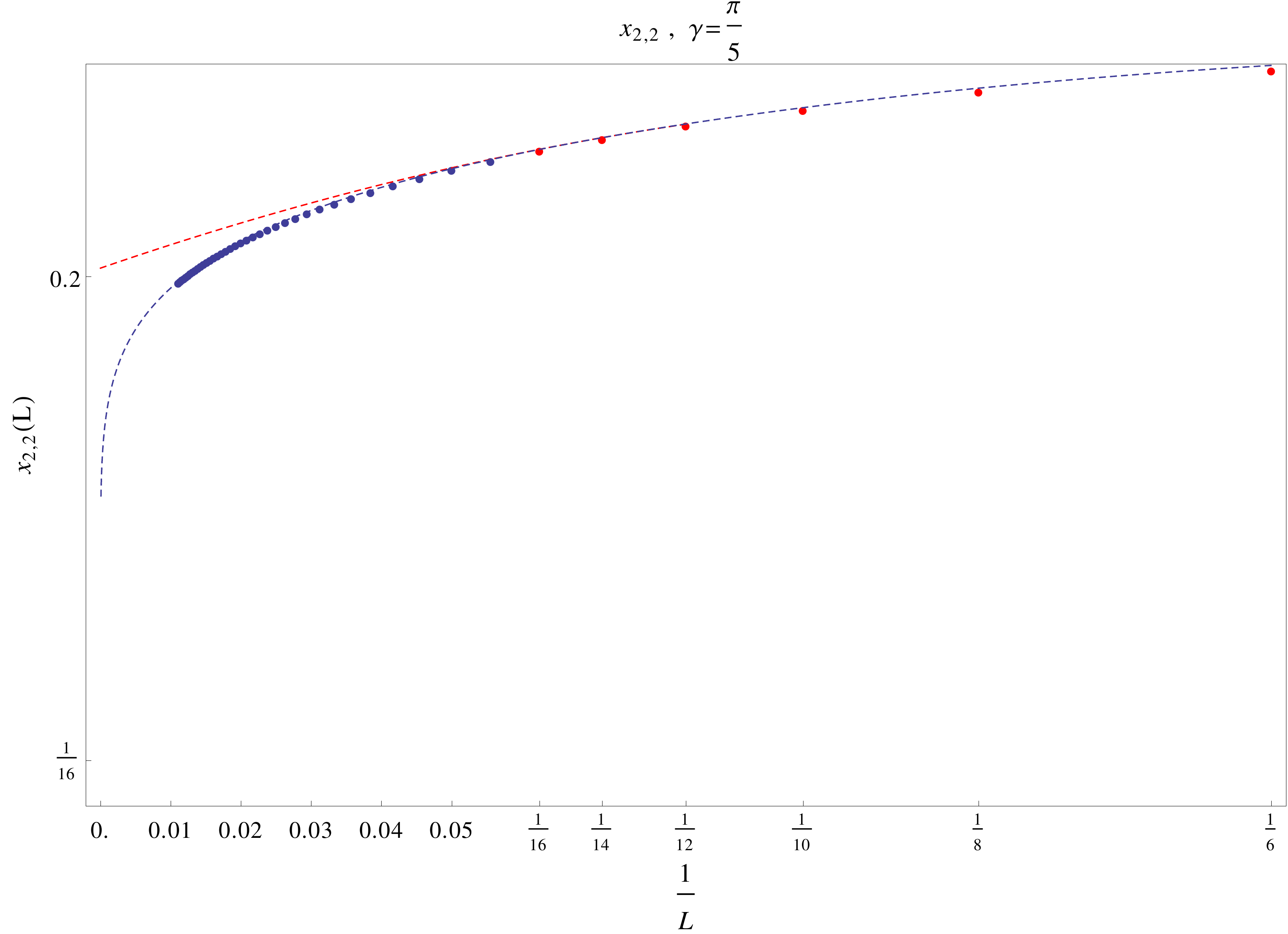}
 \end{center}
 \caption{Finite-size estimates of the exponent $x_{2,2}$ for $k=3$, plotted against $1/L$.
 The red points are the data for $L \leq 16$ that were found in \cite{FJ08} by direct
 diagonalisation of the transfer matrix. The blue points, extending this to $L \le 90$, were
 obtained here by numerical solution of the Bethe Ansatz equations. The extrapolation
 to $L \to \infty$ that led \cite{FJ08} to the erroneous value $x_{2,2} = 0.200(2)$ is shown
 as a red dashed curve. Our present extrapolation (blue dashed curves) takes into
 account logarithmic corrections to scaling and leads to $x_{2,2} = \frac{1}{16}$.
 See the main text for details.}
 \label{fig:x22scaling}
\end{figure}

Consider as an example the exponent $x_{2,2}$ for $k=3$. The finite-size estimates for $L \le 90$ are
plotted against $1/L$ in figure~\ref{fig:x22scaling}. Using direct diagonalisation of the transfer
matrix, the authors of \cite{FJ08} had however
only access to the range $L \le 16$, shown as red points in the figure. It appeared quite
reasonable to extrapolate to the $L \to \infty$ limit by fitting the last few points to a 
second order polynomial in $1/L$. This technique usually gives very accurate results
--- even in difficult situations \cite{VJ_FK_spin} --- and provides what appears to be
a quite reasonable fit to the red data points. It leads to the estimate $x_{2,2} = 0.200(2)$
given in Ref.~\cite{FJ08}.
The data for larger sizes (blue points) however makes it evident that this extrapolation
is incorrect. A much better fit is obtained by taking into account logarithmic corrections
as in (\ref{eq:ciuntwisted}). The blue dashed curve shows the form
$x_{2,2}(L) = \frac{1}{16} + \frac{A}{\left(B+\log L\right)^2}$, with
$A \simeq 28.95$ and $B \simeq 10.05$, in agreement with the asymptotic value
$x_{2,2} = \frac{1}{4(k+1)} = \frac{1}{16}$ of (\ref{watermelon_both_nonzero}).

Alternatively one may proceed without fixing $x_{2,2} = \frac{1}{16}$ as follows.
Using three successive sizes $(L,L+2,L+4)$ we first fit to the form
$x_{2,2}(L) = x_{2,2}^{\rm ext}(L) + \frac{A}{\left(B+\log L\right)^2}$
to obtain a series of extrapolants $x_{2,2}^{\rm ext}(L)$. Plotting those against
$1/L$ we observe a residual finite-size dependence which is almost linear for $L \ge 50$.
Fitting therefore $x_{2,2}^{\rm ext}(L) = x_{2,2} + C/L$ we obtain finally $x_{2,2} = 0.071(6)$. This is again
in good agreement with the proposed exact value $x_{2,2} = \frac{1}{16}$.

We should stress that the conjecture (\ref{watermelon_both_nonzero}) for the watermelon
exponents ultimately comes from (\ref{cnn0conj}) depending on both twist angles.
Obviously the determination of this function of two parameters is much less sensitive to
numerical error bars than is its specialisation to particular values of $(\phi_1,\phi_2)$. 
What is however determining in this problem is that in a whole range of $(\phi_1,\phi_2)$ the corresponding state becomes discrete and the error bars intrinsically small. The unambiguous determination of the effective central charge in this region then makes really easy to conjecture its expression in the continuum.

We note in passing that the central charge (\ref{eq:measurec}) involves the state $(n_1,n_2,j)=(0,0,0)$ 
for which the twist $\phi_1 + \phi_2 = 2 \gamma$ is greater than the threshold $\gamma$ given in
(\ref{c00iconj})--(\ref{cnn0conj}), meaning that the corresponding state is discrete. Logarithmic terms
are therefore absent, explaining why the central charge could be measured precisely from finite size scaling in \cite{FJ08} (see e.g.\ Figure 5 in that reference). 

\subsection{The watermelon exponents $x_{l_1,0}$}
\label{Section:xl10}
 
Turning to the watermelon exponents $x_{l_1 \neq 0 , 0}$ (or equivalently $x_{0, l_2 \neq 0}$), special care has to be taken to locate the central charge $c_{l_1,0}$ of the original loop model within the spectrum of the vertex model. According to section \ref{section:choiceoftwists}, the twist angles that must be taken in the vertex model are now $\phi_1=0$ and $\phi_2 = \gamma$.

To this end we first consider the spectra of both models obtained at small sizes by direct diagonalisation of the transfer matrix. The results for the 25 (resp.\ 50) lowest-lying levels of the loop model (resp.\ twisted vertex model) obtained at size $L=8$ for $l_1 = 2,4,6$ are displayed in figure \ref{fig:eigenvloopsL8s2460} in appendix. Each level in the loop model is also present in the twisted
vertex model, but the vertex model contains many levels that do not occur in the loop model.
This is in agreement with the discussion in section~\ref{sec:enlarged}. More precisely, we
observe the following features:
\begin{enumerate}
 \item For $k>2$ the ground state in the vertex model is not present in the loop model.
 \item The ground state in the loop model for small $k \gtrsim 2$ corresponds initially to the first
 excited state in the vertex model, and then to an increasingy excited state upon increasing $k$.
 At a certain value $k_0$ it joins with another level in the loop model so as to form a complex
 conjugate pair. (In the figure we have $k_0 \approx 3.1$ for $l_1=2$, $k_0 \approx 4.4$ for
 $l_1=4$, and $k_0 \approx 6.1$ for $l_1=6$.)
 \item The ground state in the loop model for large $k \gg 2$ corresponds to a highly excited
 state in the vertex model. Its analytic continuation to smaller $k$ undergoes a number of
 level crossings, and eventually becomes close to the ground state again for $k \gtrsim 2$.
\end{enumerate}

To arrive at analytical expressions for the watermelon exponents of the loop model, we need
to establish a detailed understanding of the crossovers undergone by the ground state in the
original loop model, in each sector $(l_1,0)$, as $\gamma$ runs through the interval $\left[0,{\pi \over 4}\right]$. A systematic comparison between the original and enlarged (alias vertex) models for
small values of $L$ establishes the following relations:
\begin{itemize}
 \item For $\gamma$ large enough, the ground state of the loop model with $(l_1=2n_1,0)$ legs coincides with the $(n_1,0)$ ground state of the vertex model with twist angles $\phi_1 =0$ and $\phi_2=\pi-\gamma$. This observation brings along two remarks:
 \begin{enumerate}
  \item The twist $\phi_2 = \pi - \gamma$, instead of the expected $\phi_2 =\gamma$, amounts to changing the sign of the weight $2 \cos \phi_2$ of non contractible loops of colour 2. Suppose that the boundary conditions on the system are such that there is an even number $M$ of rows. Then it can be shown \cite{FSZ87} that the number of non contractible loops in each Potts model is even. It follows that the partition function (or, more precisely, the modified partition function $Z_{l_1,0}$ conditioned to supporting the given number of through-lines) is invariant under the sign change. Decomposing
  $Z_{l_1,0}$ over the transfer matrix eigenvalues $\Lambda_j$ (see e.g.\ \cite{Richard}) gives a (weighted) sum over terms of the type $(\Lambda_j)^M$, implying that the change of $\phi_2$ only has the effect of changing the sign of a subset of the $\Lambda_j$. This effect goes away upon changing to a formulation in which the transfer matrix adds two rows to the system, and hence does not alter the physics of the problem.
  \item We {\em define} the ground state of the twisted vertex model at a given twist to be the
  analytic continuation (upon gradually increasing the twist) of the state which has the lowest
  energy in the untwisted case. In other words, this is the state $(n_1,0,0)$, with effective central charge $c_{n_1,0,0}(0,\pi - \gamma)$, using the notation established above. Because of the
  level crossings apparent in figure~\ref{fig:eigenvloopsL8s2460}, combined with the fact that
  the desired twist ($\phi_1=0$ and $\phi_2=\pi-\gamma$) depends on $\gamma$, 
  it might very well be that for some
  values of $\gamma$ the ground state of the twisted vertex model (thus defined) is not the
  lowest energy state.
  \end{enumerate}
 \item For small $\gamma$, the ground state of the loop model corresponds to some excited state of the vertex model at twists $\phi_1 =0$, $\phi_2=\gamma$. 
 Since this excited state does not belong to the general set of excitations $(n_1,0,j)$ we studied so far, we choose to call it $(n_1,0,0)'$, and write the corresponding central charge $c'_{n_1,0,0}(\phi_1,\phi_2)$. When $n_1$ is odd this state is actually a doublet of states, with respective momenta $\pm 1$ (defined with respect to the one site translation operator in the untwisted case). From the point of view of Bethe roots, the states in this doublet are anticonjugate to each other and we will need to consider only one of the two, say that of momentum $+1$, which we will still call $(n_1,0,0)'$.
\end{itemize}

Our work program is therefore the following: 1) Understand the roots pattern corresponding to $(n_1,0,0)'$ and find general conjectures for $c'_{n_1,0,0}(\phi_1,\phi_2)$, and 2) then compare 
$c_{n_1,0,0}(0,\pi - \gamma)$ and $c'_{n_1,0,0}(0,\gamma)$, yielding the location of the crossover in the $L \to \infty$ limit and the watermelons $x_{2 n_1,0}$ throughout the whole regime.

\subsubsection{Roots structure of the states $(n_1,0,0)'$}

We studied the roots configuration associated with the states $(n_1,0,0)'$ for $n_1=1,2,3$, which are represented in appendix \ref{app:RootsConfigs}. These configurations turn out to have the same qualitative structure as those of the states $(n_1,0,0)$ at large $\phi_1$, and differ only by their (properly defined) set of Bethe integers.
 Moreover, these seem to undergo no qualitative change as arbitrary values of the twist angles are taken --- at least we can affirm they do not in the whole region where we computed the corresponding effective central charges (see the next paragraph), which is enough for our purpose here.

\subsubsection{Central charges $c'_{n_1,0,0}$}

\begin{figure}
\begin{center}
 \includegraphics[scale=0.6]{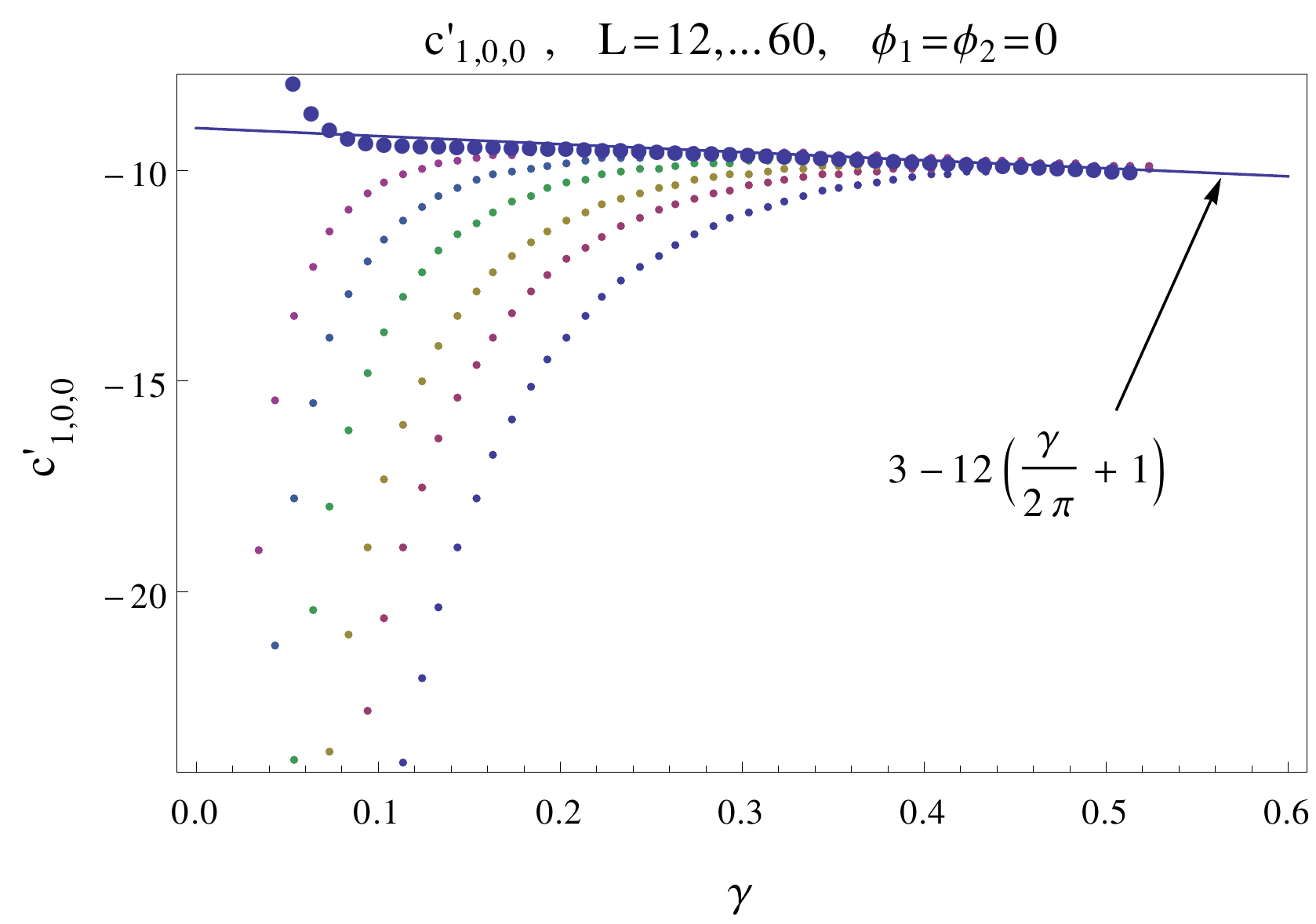}
 \end{center}
\caption{Effective central charge of the state $(1,0,0)'$ in the $n_1 = 1, n_2 =0$ sector at zero twist, plotted as a function of $\gamma$ for various sizes.
The thick blue dots correspond to an $L \to \infty$ extrapolation, and we plotted in comparison the conjectured expression.}                                 
\label{fig:cprime100_gamma}                          
\end{figure}

Turning to the corresponding effective central charges, we repeat for the states $(n_1,0,0)'$ what we did for all other states, namely determine the $\gamma$ dependence of these central charges at zero twist angles, then turn to the $\phi_1, \phi_2$ dependence and possibly to different regimes.   
We show for instance in figure \ref{fig:cprime100_gamma} the measures of $c'_{1,0,0}(0,0)$ as  a function of $\gamma$, leading to the following conjecture
\begin{equation}
 c'_{1,0,0} = c_{1,0,0} - 12 
 \label{eq:cprime1=c1-12}
\end{equation}

More generally, we found using results for sizes up to $L=100$ support (very good at $n_1=1,2$, but less perfect at $n_1=3$ where sizes $L>10$ could not be studied) for the following conjecture 
\begin{equation}
 c'_{n_1,0,0} \left(\phi_1, \phi_2 \right) = \begin{cases}  3-12 n_1 -6 {\left(n_1\right)^2 \gamma \over \pi}  - 6 \frac{\left(\phi_1\right)^2 + \left(\phi_2\right)^2}{\pi \gamma} +o(1) & \mbox{for } \phi_1 + \phi_2 \leq \left(|n_1|-1\right)\gamma \,,
\\  3-12 n_1 - 6 {\left(n_1\right)^2\gamma \over \pi} - 6 \frac{\left(\phi_1\right)^2 + \left(\phi_2\right)^2}{\pi \gamma} + 3 \frac{\left(\phi_1 + \phi_2-\left(|n_1|-1\right)\gamma\right)^2}{\gamma(\pi - \gamma)} & \mbox{for } \phi_1 + \phi_2 \geq \left(|n_1|-1\right)\gamma \,.
\end{cases}
 \label{cn00conj}
\end{equation}
This expression coincides with (\ref{cnn0conj}) for $n_2 = 0$, thus lending further credibility
to the conjecture, and has the now familiar structure in terms of discrete states.
This implies that the states $(n_1,0,0)'$ are part of a continuum for small values of the twist parameters, and therefore exhibit logarithmic terms (shown by the $o(1)$ notation in the first
line) of the same nature as those of (\ref{eq:ciuntwisted}).

\subsection{Conclusion: the watermelon exponents $x_{2 n_1,0}$}

We are now ready to proceed to the last step of our program concerning the calculation of the exponents $x_{2 n_1,0}$. 
A explained all through this section, regime III is the stage of a crossover between two states, for which we found the following effective central charges 
\begin{equation}
 c_{n_1,0,0} \left(0,\pi - \gamma\right) = \begin{cases}  3- 6 {\left(n_1\right)^2 \gamma \over \pi}  - 6 \frac{\left(\pi - \gamma\right)^2 }{\pi \gamma} +o(1) & \mbox{for } n_1 \geq k \,,
\\  3 - 6 {\left(n_1\right)^2\gamma \over \pi} - 6 \frac{\left(\pi - \gamma\right)^2 }{\pi \gamma} + 3 \frac{\left(\pi - \left(|n_1|+2\right)\gamma\right)^2}{\gamma(\pi - \gamma)} & \mbox{for } n_1 \leq k \,,
\end{cases}
\label{eq:cgamma0}
\end{equation}
and
\begin{equation}
c'_{n_1,0,0}\left(0, \gamma \right)  = \begin{cases}  3 - 6\frac{\left(n_1\right)^2\gamma}{\pi} - 12 n_1 - 6 \frac{\gamma}{\pi} + o(1)
& \mbox{for } n_1 \geq  2 \,,
\\ 3 -  6\frac{\left(n_1\right)^2\gamma}{\pi} - 12 n_1 - 6 \frac{\gamma}{\pi}
+ 3 \frac{\left(n_1-2\right)^2}{\pi -\gamma} & \mbox{for } n_1 \leq 2 \,.
\end{cases}
\label{eq:cprimegamma0}
\end{equation}

Let us treat as a warm-up the particular case of $x_{2,0}$, that is $n_1=1$. 
For this case only the second line in (\ref{eq:cprimegamma0}) is relevant. The same is true for the second line in (\ref{eq:cgamma0}), for any value of $\gamma$ in regime III (i.e., $k \ge 2$).
One easily sees that in the whole regime the former central charge is larger, as a matter of fact the two meet at $\gamma = {\pi \over 4}$. We therefore expect that for sufficiently large sizes the exponent $x_{2,0}$ be governed by $c'_{1,0,0}$ throughout regime III (or, in other words, the crossover in the limit $L \to \infty$ is located at $\gamma = {\pi \over 4}$). This leads to
\begin{equation}
 x_{2,0} = - \frac{c'_{1,0,0}(0,\gamma) - c_{0,0,0}(\gamma,\gamma)}{12} = 1 \,
\end{equation}
independently of $\gamma$. In this case we therefore confirm a conjecture made in \cite{FJ08}.
Note that this exponent was measured with very good precision in the latter reference, the reason being, as discussed for the central charge in the end of section \ref{section:cxnn}, that at twist $\phi_1=0, \phi_2=\gamma$ the state $(1,0,0)'$ is a discrete state, and hence does not possess any disturbing logarithmic terms its corresponding finite-size effective central charge.

Now for $n_1 \geq 2$ we can always use the first line in (\ref{eq:cprimegamma0}), which is the one relevant for describing the twists $\phi_1=0, \phi_2=\pi-\gamma$. Let us define a ``critical'' value of $k$ by
\begin{equation}
 k_c(n_1) \equiv \sqrt{2 \left(n_1\right)^2+2 n_1+1}+n_1-1 \,.
 \label{defkc}
\end{equation}
There are then three regimes:
\begin{itemize}
 \item $2 \leq k \leq n_1$: Here $c_{n_1,0,0}$ is described by the first line of (\ref{eq:cgamma0}), and is bigger than $c'_{n_1,0,0}$. 
 The corresponding watermelon exponent reads 
 \begin{equation}
  x_{2 n_1,0} =  - \frac{c_{1,0,0}(0,\pi-\gamma) - c_{0,0,0}(\gamma,\gamma)}{12}  = \frac{2 (k+1) \left(n_1\right)^2 +k (2 k (k+3)+3)}{4 (k+1) (k+2)} \,.
 \end{equation}
 \item $n_1 \leq k \leq k_c(n_1)$: Here $c_{n_1,0,0}$ is described by the second line of (\ref{eq:cgamma0}), and is bigger than $c'_{n_1,0,0}$. 
  The corresponding watermelon exponent is
 \begin{equation}
  x_{2 n_1,0}=  - \frac{c_{1,0,0}(0,\pi-\gamma) - c_{0,0,0}(\gamma,\gamma)}{12}  =   \frac{k \left(k^2+2 k (n_1+2)+\left(n_1\right)^2+4 n_1+3\right)}{4 (k+1) (k+2)} \,.
 \end{equation}
 \item $k \geq k_c(n_1)$: Now $c'_{n_1,0,0}$, described by the first line in (\ref{eq:cprimegamma0}), is  bigger than $c_{n_1,0,0}$.
  The corresponding watermelon exponent is in this case
 \begin{equation}
  x_{2 n_1,0} =  - \frac{c'_{1,0,0}(0,\gamma) - c_{0,0,0}(\gamma,\gamma)}{12} = -\frac{k}{4 \left(k^2+3 k+2\right)}+\frac{\left(n_1\right)^2}{2 k+4}+n_1 \,.
 \end{equation}
\end{itemize}

Note that the location $k_c$ of the transition between the regimes dominated by
$c_{n_1,0,0}$ and $c'_{n_1,0,0}$ is an increasing function of $n_1$, in agreement with
numerical results (such as those shown in figure~\ref{fig:eigenvloopsL8s2460})
that we have obtained by directly diagonalising the relevant transfer matrices for small sizes $L$.

Our result for $x_{2 n_1,0}$ with $n_1 \ge 2$ can be compared with the numerical results
given in Table 1 of \cite{FJ08}. Note that the latter have relatively large error bars, and in
some cases did not converge.

\section{Conclusion and discussion}
\label{section:conclusion}

We have studied an integrable case of two coupled $Q$-state antiferromagnetic Potts models on the square lattice,
first discovered by Martins and Nienhuis \cite{MartinsNienhuis} and further
studied by Fendley and Jacobsen \cite{FJ08},
who identified its continuum limit from an argument of level-rank duality.
Going beyond this, we have exhibited an
exact mapping of the lattice model to the $a_3^{(2)}$ vertex model \cite{GalleasMartins} in regime III,
which corresponds to the range $k \in [2,\infty)$ in the parameterisation (\ref{eq:loopweight}).

This mapping has allowed us in particular to extract the Bethe Ansatz equations
(see also \cite{MartinsNienhuis}).
Studying these numerically, and drawing on
formal analogies with the $a_2^{(2)}$ model \cite{VJS:a22}, we have established that the
continuum limit contains two compact bosons and one non compact boson. The non compact
degree of freedom entails a continuous spectrum of critical exponents. When twisting
the vertex model discrete states emerge from the continuum at particular twist angles that we have precisely identified.

Results on the coupled Potts models --- in their formulation as a dense two-colour loop model \cite{FJ08} --- then followed from identifying the twist angles that ensure the equivalence between the vertex model and the loop model in various sectors. For the ground state sector we recovered
the central charge 
\begin{equation}
 c = {3 k^2 \over (k+1)(k+2)} \,,
 \label{c_again}
\end{equation}
in agreement with \cite{FJ08}. Improving on this we have also given complete results for
the magnetic exponents of the watermelon type, corresponding to imposing given numbers
$(2n_1,2n_2)$ of propagating through-lines in each Potts model.
The watermelon exponents $x_{2n_1, 2n_2}$, in the case where both $n_1$ and $n_2$ are non zero,
were found to be
\begin{equation}
  x_{2 n_1, 2 n_2} = \frac{\left(n_1\right)^2 + \left(n_2\right)^2-2}{2}\frac{1}{k+2} + {1 \over  4(k +1)} . 
\end{equation}
The exponents $x_{2 n_1,0}$ for $n_1 \neq 0$ (or equivalently, $x_{0,2 n_2}$ with $n_2 \neq 0$)
exhibit a delicate crossover when $k$ runs through the interval $[2,\infty)$. Aside from the particular case $x_{2,0}=1$, we find for $n_1 \ge 2$ three different regimes:
\begin{equation}
 x_{2 n_1,0} = \begin{cases}
 \frac{2 (k+1) \left(n_1\right)^2 +k (2 k (k+3)+3)}{4 (k+1) (k+2)} &
 \mbox{for } 2 \leq k \leq n_1 \,, \\
 \frac{k \left(k^2+2 k (n_1+2)+\left(n_1\right)^2+4 n_1+3\right)}{4 (k+1) (k+2)} &
 \mbox{for } n_1 \leq k \leq k_c(n_1) \,, \\
 -\frac{k}{4 \left(k^2+3 k+2\right)}+\frac{\left(n_1\right)^2}{2 k+4}+n_1 &
 \mbox{for } k \geq k_c(n_1) \,,
\end{cases}
\end{equation}
where $k_c(n_1)$ is defined in (\ref{defkc}).

The numerical study made in \cite{FJ08} for the first few watermelon exponents was based
on the numerical diagonalisation of the transfer matrix for sizes ranging up to $L=16$.
The strong logarithmic corrections, produced by the non compact nature of the continuum limit,
were unfortunately unknown to the authors of \cite{FJ08} and led them to numerical estimates
(and a couple of conjectures) that --- with the exception of $c$ and $x_{2,0}$ for which no
logarithms are present --- have been invalidated by the present work
(see section~\ref{FJcomp} for a detailed comparison of the numerical analyses).

This might be a useful
lesson for the future, since we now know at least three different loop models \cite{IkhlefJS3,VJS:a22}
having a non compact continuum limit.

We should point out that it follows from an argument in \cite{DJLP99} that when $Q=3$ (i.e., $k=4$) 
the two Potts models decouple. More precisely, the free energies in the ground state sector
for the dense two-colour loop model (with $\lambda=\lambda_c$) and the decoupled models (with $\lambda=1$) are related by
\begin{equation}
 f(\lambda_c,L) = f(1,L) + \frac12 \log \left(2 (2 + \sqrt{3}) \right) \qquad \mbox{(for $Q=3$)} \,.
\end{equation}
Consequently the central charge (\ref{c_again}) is $c = \frac85 = 2 \times \frac45$.
However, this does not imply that the spectrum of excitations at $k=4$ is also related to
that of a single $3$-state Potts model. In particular, the watermelon operators given above
are manifestly completely different. Moreover, we stress that the $k=4$ model will also have
non compact excitations --- a feature which is obviously absent in the $3$-state Potts model.%
\footnote{A similar remark can be made about the relation between the square-lattice Ising model
at its ferromagnetic and antiferromagnetic critical points. In the ground state sector these
two models are related by a well-known mapping (change the sign of the coupling constant
and flip the spins on one sublattice), and both have $c=\frac12$. However, the loop model
underlying the $Q=2$ state antiferromagnetic Potts model \cite{JS_AFPotts} has an excitation
spectrum that is quite different from that of its ferromagnetic counterpart.}

We have left several directions for future work. For example, it would be interesting to compute
also the energy-like critical exponents of the coupled Potts models. Another open issue is to
study non-periodic boundary conditions, and to consider in particular boundary extensions of the
underlying Temperley-Lieb algebra along the lines of \cite{confbound}. On the integrability
side, setting up the non-linear integral equations (NLIE) might allow one to actually prove our
formulae for the critical exponents and for the density of states. A field-theoretical formulation
of the $a_3^{(2)}$ model will appear elsewhere in the more general $a_n^{(2)}$ setting \cite{VJS:an2}.

Let us finally mention that the dense two-colour loop model studied here has a dilute counterpart
which is related to a truncated version of the plateau transition in the integer quantum Hall effect
\cite{IkhlefFC}. The dilute model is expected to contain non compact features as well, and we hope to
report more on this issue elsewhere.

\subsection*{Acknowledgments}

We thank Paul Fendley for comments on the manuscript.
This work was supported by the French Agence Nationale pour la Recherche
(ANR Projet 2010 Blanc SIMI 4: DIME) and the Institut Universitaire de France (JLJ).

\appendix

\section{Low-lying levels of the original and enlarged loop models}
\label{app:A}

\begin{figure}
\begin{center}
 \includegraphics[scale=0.4]{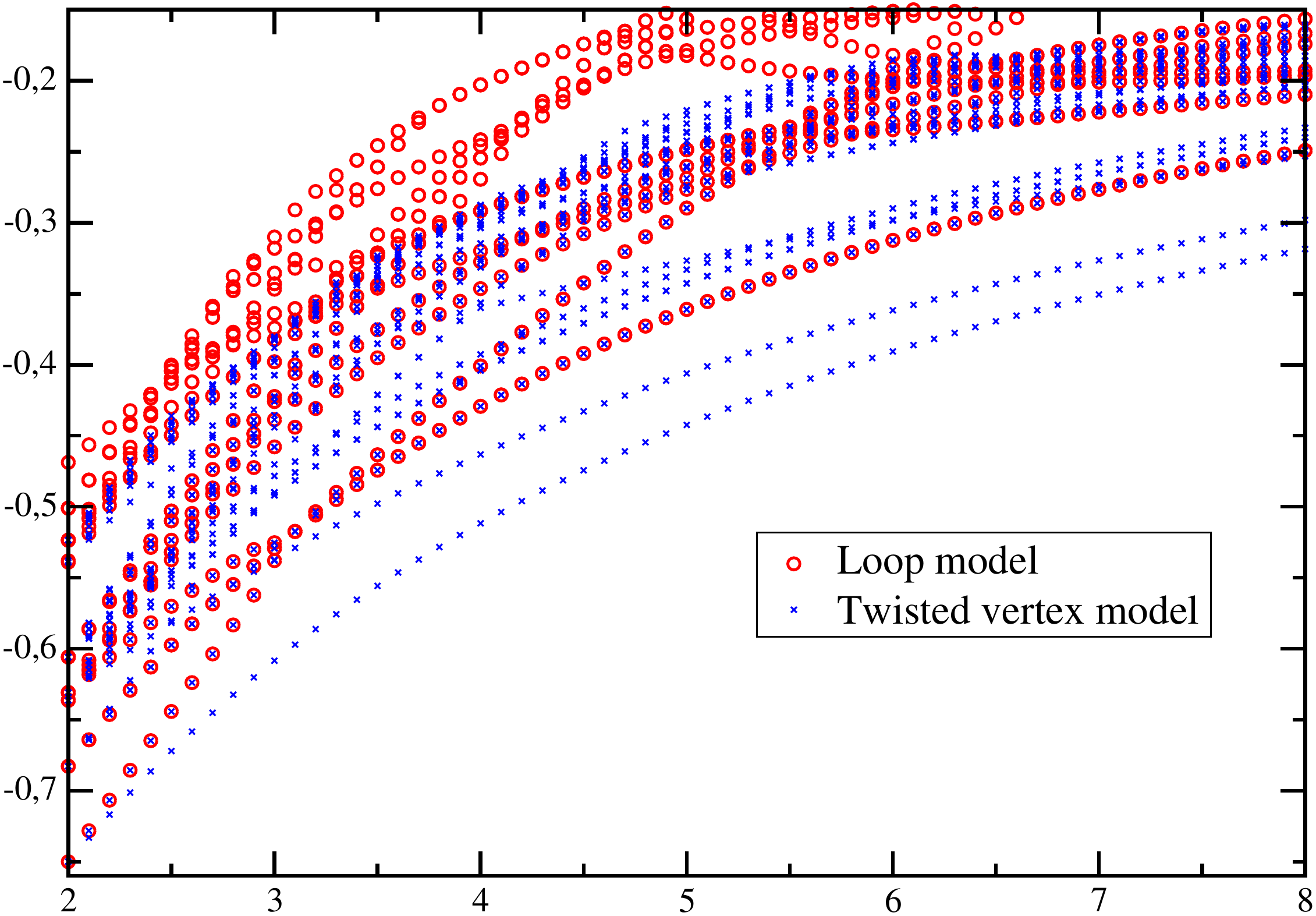}
  \includegraphics[scale=0.4]{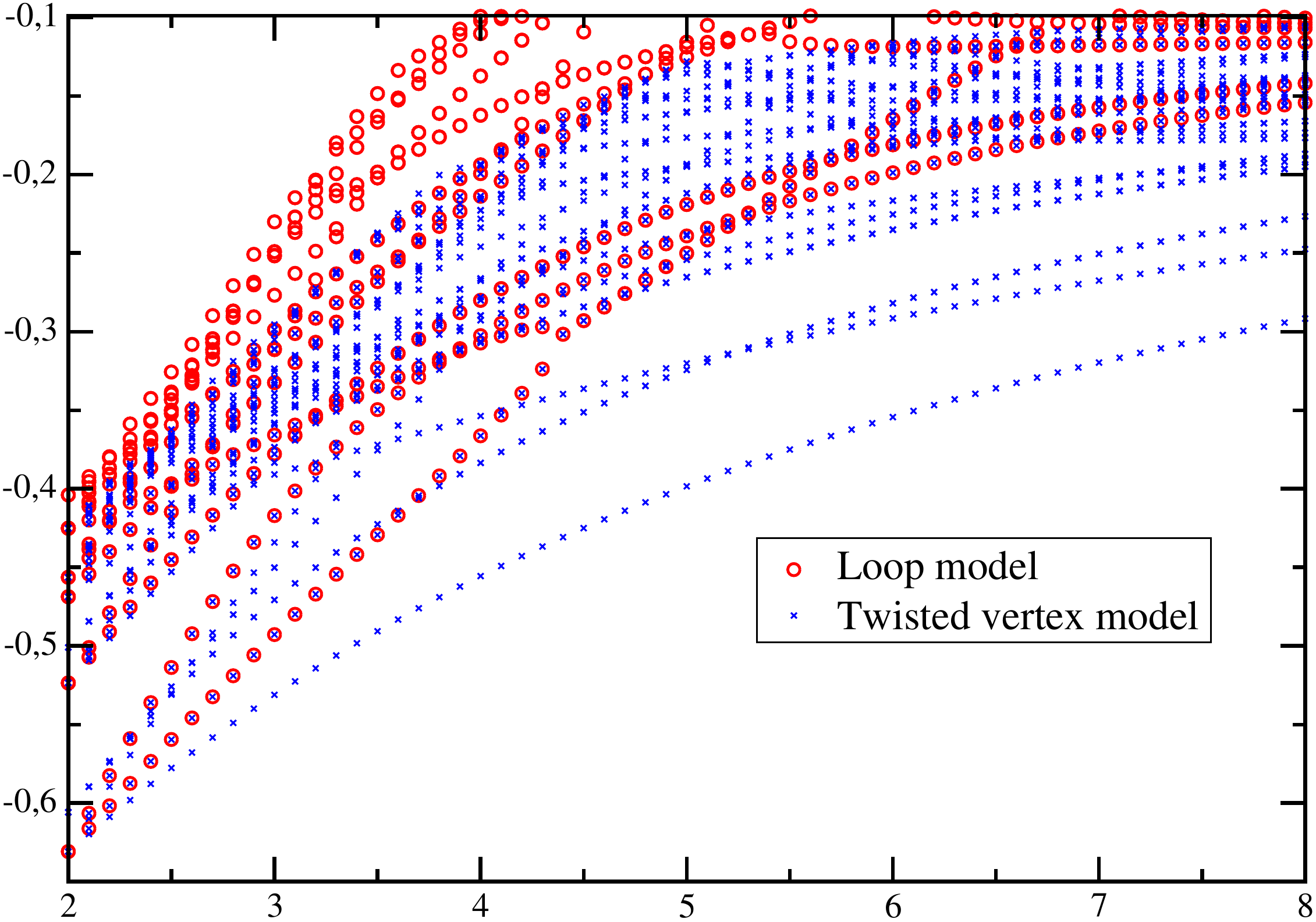}
   \includegraphics[scale=0.4]{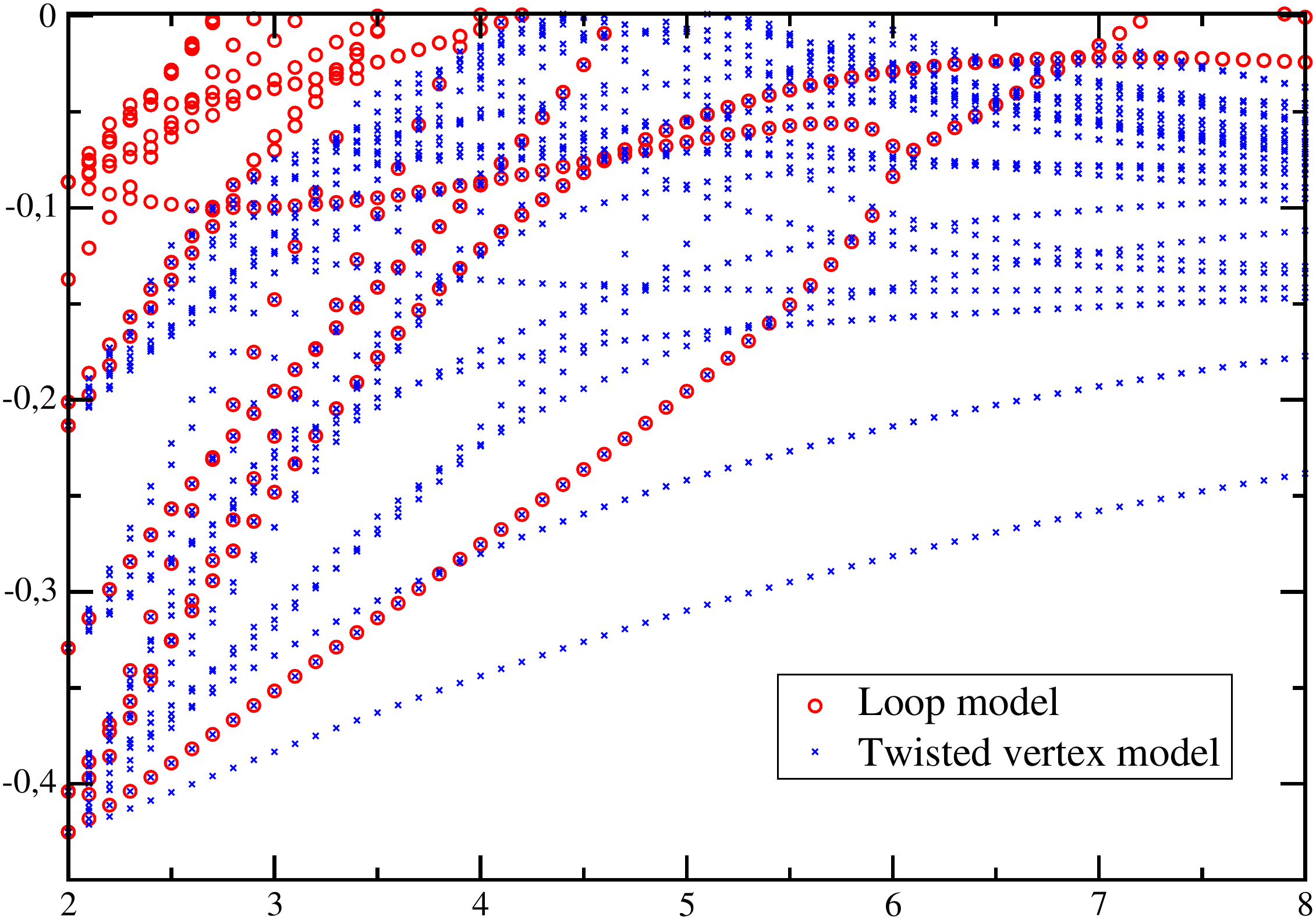}
 \end{center}
\caption{Low-lying levels of the original (red circles) and enlarged (blue crosses) loop models measured at size $L=8$ in sector $(l_1,l_2)=(2,0), (4,0), (6,0)$ from top to bottom. We show the free energies per site as functions of $k$. All the levels in the loop model are also in the twisted vertex model, but the latter contains many levels not present in the loop model. In particular, the ground state of the loop model does not coincide with that of the twisted vertex model.
Moreover, the ground state for large $k$ is not the analytic continuation of that for small $k$, 
due to crossover phenomena which are detailed in the main text.
}
\label{fig:eigenvloopsL8s2460}                          
\end{figure}

On figure \ref{fig:eigenvloopsL8s2460} we display the low-lying levels obtained from direct diagonalization of the original and enlarged loop models. The main conclusions are drawn in section \ref{Section:xl10}.

\section{Roots configurations at finite size}
\label{app:RootsConfigs}

In this appendix we display the roots configurations associated with some of the levels mentioned in the main text. 

\begin{itemize}

\item On figure \ref{fig:E200g030L16} we reproduce the roots configuration associated with the state $(2,0,0)$, namely the ground state of the sector $(n_1,n_2)=(2,0)$ for $L=16$, $\gamma =\frac{3}{10}$. 
\begin{figure}
\begin{center}
 \includegraphics[width=90mm,height=70mm]{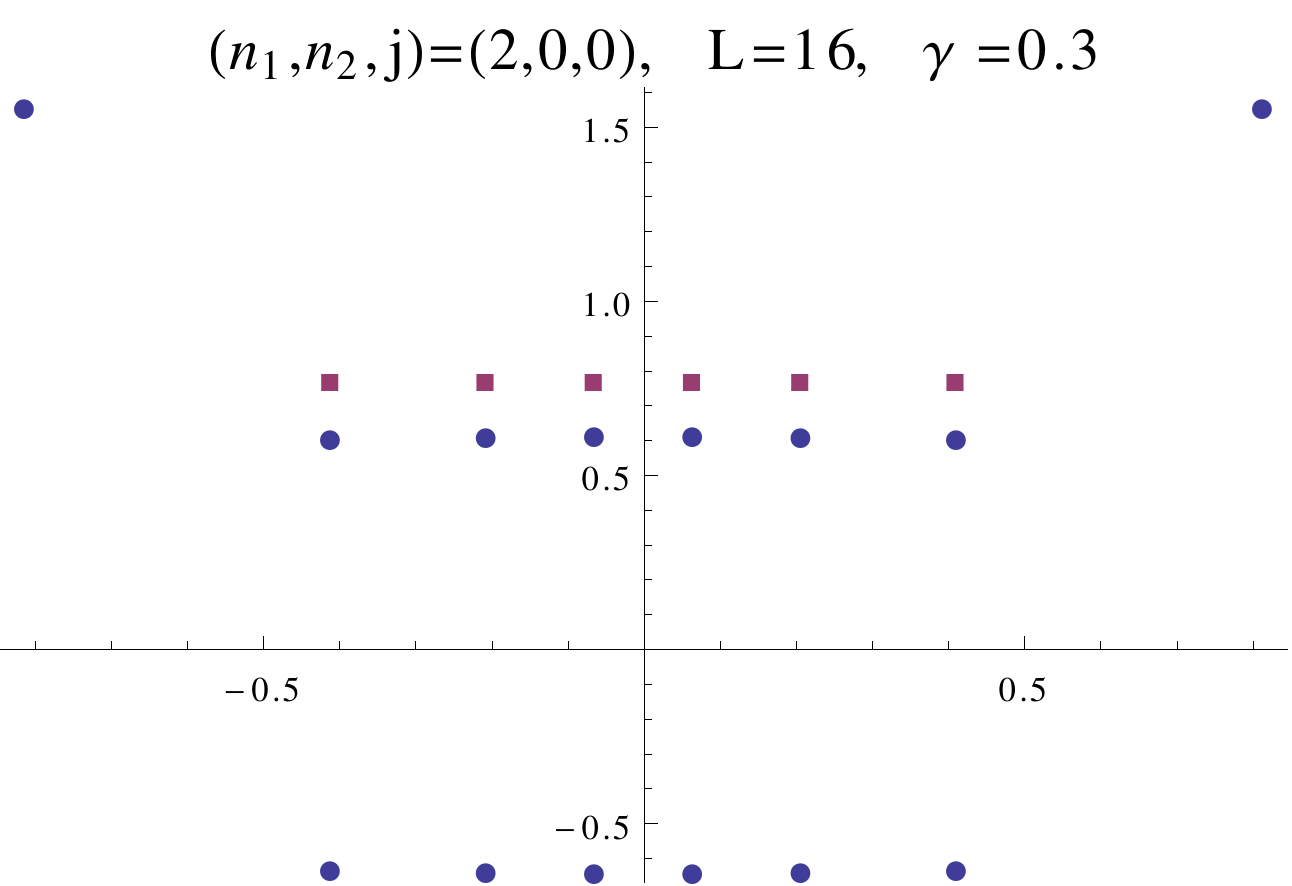}
 \end{center}
\caption{Configuration of the $\lambda$ (in blue) and $\mu$ (in purple) roots corresponding to the ground state of regime III in the sector $(n_1,n_2) = (2,0)$, at $\gamma = \frac{3}{10}$ and for a system size $L=16$.}                                 
\label{fig:E200g030L16}                          
\end{figure}

\item On figure \ref{fig:nOOtwist}, we represent schematically the roots configurations associated with the states $(n_1,0,0)$ as a function of the twist $\phi_1$. It turns out these only depend on $\phi_1$ only, and present qualitative changes as $\phi_1$ reaches some particular integer multiples of $\gamma$ : as the value of $\phi_1$ is increased, some of the excited $\lambda$-roots (that is, those with imaginary part $\pi \over 2$) can become real or form complexes that we shall refer to as ``2*-strings'' and which are strings with imaginary part close to $\pm \gamma \over 2$.

\newcommand{\nOOtwistO}{
\begin{tikzpicture}[scale=0.5]
\draw [black, line width=0.1] (0,-0.8)  -- (0,2); 
\draw [black, line width=0.2] (-3,0) -- (3,0); 
\draw [black, dashed, line width=0.2] (-2.8,1.5) -- (3,1.5);
\draw[fill] (-3,1.6) node[scale=1] {$\pi \over 2$};
\draw[blue] (-2.,1.5) node[scale=0.75] {$\bullet$} ;
\draw[blue] (-1.3,1.5) node[scale=0.75] {$\bullet$} ;
\draw[blue] (-0.5,1.5) node[scale=0.75] {$\bullet$} ;
\draw[blue] (0.5,1.5) node[scale=0.75] {$\bullet$} ;
\draw[blue] (1.3,1.5) node[scale=0.75] {$\bullet$} ;
\draw[blue] (2.,1.5) node[scale=0.75] {$\bullet$} ;
\draw[blue, line width=0.3mm, rounded corners=10pt] (-1.5,0.5) -- (0,0.8) -- (1.5,0.5);
\draw[blue, line width=0.3mm, rounded corners=10pt] (-1.5,-0.5) -- (0,-0.8) -- (1.5,-0.5);
\draw [decorate,decoration={brace,amplitude=5pt},xshift=0pt,yshift=-2pt]
(1.5,-0.8) -- (-1.5,-0.8)node [black,midway,yshift=-10pt] {\footnotesize{${L \over 2} - n_1$ 2-strings}};
\draw [decorate,decoration={brace,amplitude=5pt},xshift=0pt,yshift=2pt]
(-2,1.5) -- (2,1.5)node [black,midway,yshift=10pt] {\footnotesize{$n_1$}};
\end{tikzpicture}}

\newcommand{\nOOtwistOg}{
\begin{tikzpicture}[scale=0.5]
\draw [black, line width=0.1] (-0.5,-0.8)  -- (-0.5,2); 
\draw [black, line width=0.2] (-3,0) -- (3,0); 
\draw [black, dashed, line width=0.2] (-3,1.5) -- (3,1.5);
\draw[blue] (-2.,1.5) node[scale=0.75] {$\bullet$} ;
\draw[blue] (-1.3,1.5) node[scale=0.75] {$\bullet$} ;
\draw[blue] (-0.5,1.5) node[scale=0.75] {$\bullet$} ;
\draw[blue] (0.5,1.5) node[scale=0.75] {$\bullet$} ;
\draw[blue] (1.3,1.5) node[scale=0.75] {$\bullet$} ;
\draw[blue] (2.2,1.5) node[scale=0.75] {$\bullet$} ;
\draw[blue, line width=0.3mm, rounded corners=10pt] (-1.5,0.5) -- (0,0.8) -- (1.5,0.5);
\draw[blue, line width=0.3mm, rounded corners=10pt] (-1.5,-0.5) -- (0,-0.8) -- (1.5,-0.5);
\end{tikzpicture}}

\newcommand{\nOOtwistg}{
\begin{tikzpicture}[scale=0.5]
\draw [black, line width=0.1] (-0.5,-0.8)  -- (-0.5,2); 
\draw [black, line width=0.2] (-3,0) -- (3,0); 
\draw [black, dashed, line width=0.2] (-3,1.5) -- (3,1.5);
\draw[blue] (-2.,1.5) node[scale=0.75] {$\bullet$} ;
\draw[blue] (-1.3,1.5) node[scale=0.75] {$\bullet$} ;
\draw[blue] (-0.5,1.5) node[scale=0.75] {$\bullet$} ;
\draw[blue] (0.5,1.5) node[scale=0.75] {$\bullet$} ;
\draw[blue] (1.3,1.5) node[scale=0.75] {$\bullet$} ;
\draw[blue] (2.5,1.5) node[scale=0.75] {$\bullet$} ;
\draw (2.7,1.7) node[scale=1] {\footnotesize{$\to \infty$}} ;
\draw[blue, line width=0.3mm, rounded corners=10pt] (-1.5,0.5) -- (0,0.8) -- (1.5,0.5);
\draw[blue, line width=0.3mm, rounded corners=10pt] (-1.5,-0.5) -- (0,-0.8) -- (1.5,-0.5);
\end{tikzpicture}}

\newcommand{\nOOtwistgIIg}{
\begin{tikzpicture}[scale=0.5]
\draw [black, line width=0.1] (-0.7,-0.8)  -- (-0.7,2); 
\draw [black, line width=0.2] (-3,0) -- (3,0); 
\draw [black, dashed, line width=0.2] (-3,1.5) -- (3,1.5);
\draw[blue] (-2.,1.5) node[scale=0.75] {$\bullet$} ;
\draw[blue] (-1.3,1.5) node[scale=0.75] {$\bullet$} ;
\draw[blue] (-0.5,1.5) node[scale=0.75] {$\bullet$} ;
\draw[blue] (0.5,1.5) node[scale=0.75] {$\bullet$} ;
\draw[blue] (1.3,1.5) node[scale=0.75] {$\bullet$} ;
\draw[blue] (2.2,0) node[scale=0.75] {$\bullet$} ;
\draw[blue, line width=0.3mm, rounded corners=10pt] (-1.5,0.5) -- (0,0.8) -- (1.5,0.5);
\draw[blue, line width=0.3mm, rounded corners=10pt] (-1.5,-0.5) -- (0,-0.8) -- (1.5,-0.5);
\end{tikzpicture}}

\newcommand{\nOOtwistIIg}{
\begin{tikzpicture}[scale=0.5]
\draw [black, line width=0.1] (-1,-0.8)  -- (-1,2); 
\draw [black, line width=0.2] (-3,0) -- (3,0); 
\draw [black, dashed, line width=0.2] (-3,1.5) -- (3,1.5);
\draw[blue] (-2.,1.5) node[scale=0.75] {$\bullet$} ;
\draw[blue] (-1.3,1.5) node[scale=0.75] {$\bullet$} ;
\draw[blue] (-0.5,1.5) node[scale=0.75] {$\bullet$} ;
\draw[blue] (0.5,1.5) node[scale=0.75] {$\bullet$} ;
\draw[blue] (2.5,1.5) node[scale=0.75] {$\bullet$} ;
\draw (2.7,1.7) node[scale=1] {\footnotesize{$\to \infty$}} ;
\draw[blue] (2.5,0) node[scale=0.75] {$\bullet$} ;
\draw (2.7,0.2) node[scale=1] {\footnotesize{$\to \infty$}} ;
\draw[blue, line width=0.3mm, rounded corners=10pt] (-1.5,0.5) -- (0,0.8) -- (1.5,0.5);
\draw[blue, line width=0.3mm, rounded corners=10pt] (-1.5,-0.5) -- (0,-0.8) -- (1.5,-0.5);
\end{tikzpicture}}

\newcommand{\nOOtwistIIgIIIg}{
\begin{tikzpicture}[scale=0.5]
\draw [black, line width=0.1] (-1.5,-0.8)  -- (-1.5,2); 
\draw [black, line width=0.2] (-3,0) -- (3,0); 
\draw [black, dashed, line width=0.2] (-3,1.5) -- (3,1.5);
\draw[blue] (-2.,1.5) node[scale=0.75] {$\bullet$} ;
\draw[blue] (-1.3,1.5) node[scale=0.75] {$\bullet$} ;
\draw[blue] (-0.5,1.5) node[scale=0.75] {$\bullet$} ;
\draw[blue] (0.5,1.5) node[scale=0.75] {$\bullet$} ;
\draw[blue] (2.5,-0.25) node[scale=0.75] {$\bullet$} ;
\draw[blue] (2.5,0.25) node[scale=0.75] {$\bullet$} ;
\draw[blue, line width=0.3mm, rounded corners=10pt] (-1.5,0.5) -- (0,0.8) -- (1.5,0.5);
\draw[blue, line width=0.3mm, rounded corners=10pt] (-1.5,-0.5) -- (0,-0.8) -- (1.5,-0.5);
\end{tikzpicture}}

\newcommand{\nOOtwistngInfty}{
\begin{tikzpicture}[scale=0.5]
\draw [black, line width=0.1] (-2,-0.8)  -- (-2,2); 
\draw [black, line width=0.2] (-3,0) -- (3,0); 
\draw [black, dashed, line width=0.2] (-3,1.5) -- (3,1.5);
\draw[blue] (1.7,-0.25) node[scale=0.75] {$\bullet$} ;
\draw[blue] (1.7,0.25) node[scale=0.75] {$\bullet$} ;
\draw[blue] (2.0,-0.25) node[scale=0.75] {$\bullet$} ;
\draw[blue] (2.0,0.25) node[scale=0.75] {$\bullet$} ;
\draw[blue] (2.3,-0.25) node[scale=0.75] {$\bullet$} ;
\draw[blue] (2.3,0.25) node[scale=0.75] {$\bullet$} ;
\draw[blue] (2.8,0) node[scale=0.75] {$(\bullet)$} ;
\draw[blue, line width=0.3mm, rounded corners=10pt] (-1.5,0.5) -- (0,0.8) -- (1.5,0.5);
\draw[blue, line width=0.3mm, rounded corners=10pt] (-1.5,-0.5) -- (0,-0.8) -- (1.5,-0.5);
\end{tikzpicture}}

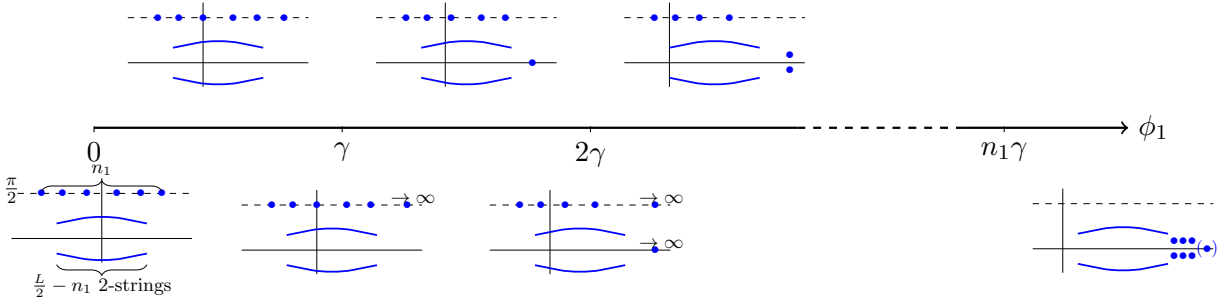
\begin{figure}
\begin{center}
\begin{tikzpicture}[scale=0.55]
\draw[black, line width=0.3mm] (0,0) -- (17,0);
\draw[black, line width=0.3mm,dashed] (17,0) -- (21,0);
\draw[black, line width=0.3mm,->] (21,0) -- (25,0);
\draw (25,0) node[right] {$\phi_1$};
\draw (0,1pt) -- (0,-3pt) node[anchor=north] {$0$};
\draw (6,1pt) -- (6,-3pt) node[anchor=north] {$\gamma$};
\draw (12,1pt) -- (12,-3pt) node[anchor=north] {$2\gamma$};
\draw (22,1pt) -- (22,-3pt) node[anchor=north] {$n_1\gamma$};
\draw (0,-2.5) node[scale=0.8] {\nOOtwistO};
\draw (3,2) node[scale=0.8] {\nOOtwistOg};
\draw (6,-2.5) node[scale=0.8] {\nOOtwistg};
\draw (9,2) node[scale=0.8] {\nOOtwistgIIg};
\draw (12,-2.5) node[scale=0.8] {\nOOtwistIIg};
\draw (15,2) node[scale=0.8] {\nOOtwistIIgIIIg};
\draw (25,-2.5) node[scale=0.8] {\nOOtwistngInfty};
\end{tikzpicture} \end{center}
\caption{Configuration of the $\lambda$-roots associated with the states $(n_1, 0,0)$ as a function of the twist $\phi_1$, for any value of the twist $\phi_2$. The $\mu$-roots are just continuously shifted on the axis of imaginary part $\pi \over 4$. 
The Fermi sea of 2-strings is represented by the curved lines, whereas the additional 2*-strings formed at large twist (with imaginary part close to $\pm \gamma \over 2$) are represented by conjugate dots. 
The single root represented between parentheses in the rightmost diagram is present only if $n_1$ is odd.
}                                 
\label{fig:nOOtwist}                          
\end{figure}

\item On figure \ref{fig:nOO1twist}, we represent schematically the roots configuration associated with the state $(0,0,1)$ as a function of the twist $\phi_1$, as an example of how the excited states in the various magnetization sectors also undergo qualitative changes. 

It is not difficult to infer the structure of the roots for any state $(n_1,n_2,j)$ from this example and the previous one, but we will not go into any more detail here. 

\newcommand{\nOOItwistO}{
\begin{tikzpicture}[scale=0.5]
\draw [black, line width=0.1] (0,-0.8)  -- (0,2); 
\draw [black, line width=0.2] (-3,0) -- (3,0); 
\draw [black, dashed, line width=0.2] (-2.8,2) -- (3,2);
\draw [black, dashed, line width=0.2] (-3,1) -- (3,1);
\draw[fill] (-3,2.1) node[scale=1] {$\pi \over 2$};
\draw[fill] (-3,1.1) node[scale=1] {$\pi \over 4$};
\draw[blue] (-2,2) node[scale=0.75] {$\bullet$} ;
\draw[blue] (2,2) node[scale=0.75] {$\bullet$} ;
\foreach \x in {0,...,15}
     		\draw[purple] (-1.4+2.8/15*\x,1) node[scale=0.75] {$\bullet$} ;
\draw[blue, line width=0.3mm, rounded corners=10pt] (-1.5,0.5) -- (0,0.8) -- (1.5,0.5);
\draw[blue, line width=0.3mm, rounded corners=10pt] (-1.5,-0.5) -- (0,-0.8) -- (1.5,-0.5);
\draw [decorate,decoration={brace,amplitude=5pt},xshift=0pt,yshift=-2pt]
(1.5,-0.8) -- (-1.5,-0.8)node [black,midway,yshift=-10pt] {\footnotesize{${L \over 2} - n_1$ 2-strings}};
\end{tikzpicture}}

\newcommand{\nOOItwistOIIg}{
\begin{tikzpicture}[scale=0.5]
\draw [black, line width=0.1] (-1,-0.8)  -- (-1,2); 
\draw [black, line width=0.2] (-3,0) -- (3,0); 
\draw [black, dashed, line width=0.2] (-2.8,2) -- (3,2);
\draw [black, dashed, line width=0.2] (-3,1) -- (3,1);
\draw[blue] (-2,2) node[scale=0.75] {$\bullet$} ;
\draw[blue] (2.1,2) node[scale=0.75] {$\bullet$} ;
\foreach \x in {0,...,14}
     		\draw[purple] (-1.4+2.8/15*\x,1) node[scale=0.75] {$\bullet$} ;
 \draw[purple] (1.9,1) node[scale=0.75] {$\bullet$} ;    		
\draw[blue, line width=0.3mm, rounded corners=10pt] (-1.5,0.5) -- (0,0.8) -- (1.5,0.5);
\draw[blue, line width=0.3mm, rounded corners=10pt] (-1.5,-0.5) -- (0,-0.8) -- (1.5,-0.5);
\end{tikzpicture}}

\newcommand{\nOOItwistIIg}{
\begin{tikzpicture}[scale=0.5]
\draw [black, line width=0.1] (-1.2,-0.8)  -- (-1.2,2); 
\draw [black, line width=0.2] (-3,0) -- (3,0); 
\draw [black, dashed, line width=0.2] (-2.8,2) -- (3,2);
\draw [black, dashed, line width=0.2] (-3,1) -- (3,1);
\draw[blue] (-1,2) node[scale=0.75] {$\bullet$} ;
\draw[blue] (2.5,2) node[scale=0.75] {$\bullet$} ;
\foreach \x in {0,...,14}
     		\draw[purple] (-1.4+2.8/15*\x,1) node[scale=0.75] {$\bullet$} ;
 \draw[purple] (2.5,1) node[scale=0.75] {$\bullet$} ;    
 \draw (2.7,1.2) node[scale=1] {\footnotesize{$\to \infty$}} ;
 \draw (2.7,2.2) node[scale=1] {\footnotesize{$\to \infty$}} ;
 \draw[blue, line width=0.3mm, rounded corners=10pt] (-1.5,0.5) -- (0,0.8) -- (1.5,0.5);
\draw[blue, line width=0.3mm, rounded corners=10pt] (-1.5,-0.5) -- (0,-0.8) -- (1.5,-0.5);
\end{tikzpicture}}

\newcommand{\nOOItwistIIgInfinity}{
\begin{tikzpicture}[scale=0.5]
\draw [black, line width=0.1] (-1.5,-0.8)  -- (-1.5,2); 
\draw [black, line width=0.2] (-3,0) -- (3,0); 
\draw [black, dashed, line width=0.2] (-2.8,2) -- (3,2);
\draw [black, dashed, line width=0.2] (-3,1) -- (3,1);
\draw[blue] (0,2) node[scale=0.75] {$\bullet$} ;
\draw[blue] (1.8,0) node[scale=0.75] {$\bullet$} ;
\foreach \x in {0,...,14}
     		\draw[purple] (-1.4+2.8/15*\x,1) node[scale=0.75] {$\bullet$} ;
 \draw[purple] (2.3,0) node[scale=0.75] {$\bullet$} ;    
 \draw[blue, line width=0.3mm, rounded corners=10pt] (-1.5,0.5) -- (0,0.8) -- (1.5,0.5);
\draw[blue, line width=0.3mm, rounded corners=10pt] (-1.5,-0.5) -- (0,-0.8) -- (1.5,-0.5);
\end{tikzpicture}}

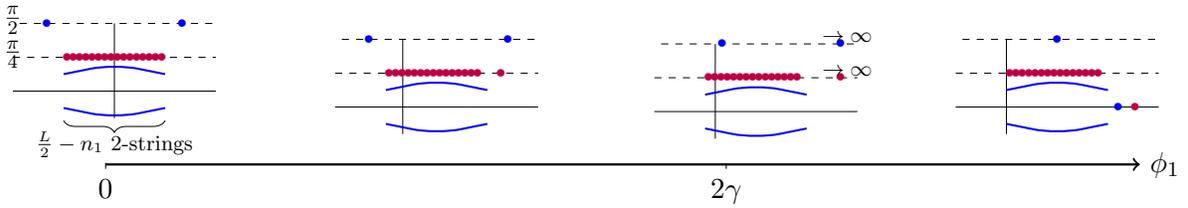
\begin{figure}
\begin{center}
\begin{tikzpicture}[scale=0.55]
\draw[black, line width=0.3mm,->] (0,0) -- (25,0);
\draw (25,0) node[right] {$\phi_1$};
\draw (0,1pt) -- (0,-3pt) node[anchor=north] {$0$};
\draw (15,1pt) -- (15,-3pt) node[anchor=north] {$2\gamma$};
\draw (0,2) node[scale=0.9] {\nOOItwistO};
\draw (8,2) node[scale=0.9] {\nOOItwistOIIg};
\draw (16,2) node[scale=0.9] {\nOOItwistIIg};
\draw (23,2) node[scale=0.9] {\nOOItwistIIgInfinity};
\end{tikzpicture} \end{center}
\caption{Configuration of the $\lambda$ (in blue) and $\mu$ (in purple) roots associated with the state $(0, 0,1)$ as a function of the twist $\phi_1$, for any value of the twist $\phi_2$. 
}                                 
\label{fig:nOO1twist}                          
\end{figure}

\item On figure \ref{fig:En00primeg030L16}, we reproduce the roots configurations found for the states $(1,0,0)'$ and $(2,0,0)'$ (for $L=16$) and the state $(3,0,0)'$ (for $L=10$) at $\gamma=\frac{3}{10}$ and zero twist. These are similar to those found for the ground states $(1,0,0)$, $(2,0,0)$, $(3,0,0)$ at large $\phi_1$ (namely, the configurations represented by the rightmost panel of figure \ref{fig:nOOtwist}), and differ only by their (properly defined) set of Bethe integers. 
 \begin{figure}
\begin{center}
 \includegraphics[width=90mm,height=70mm]{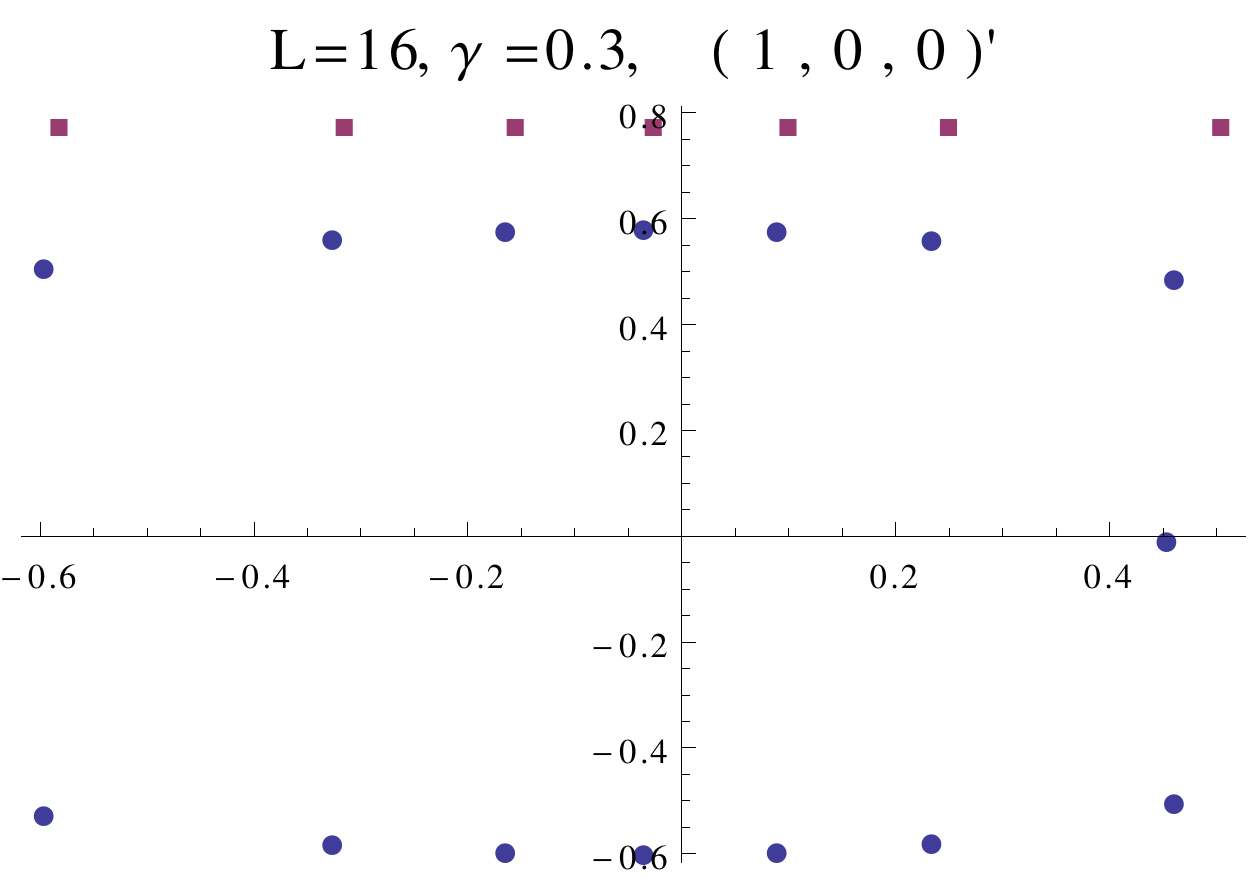}
 \\
  \includegraphics[width=90mm,height=70mm]{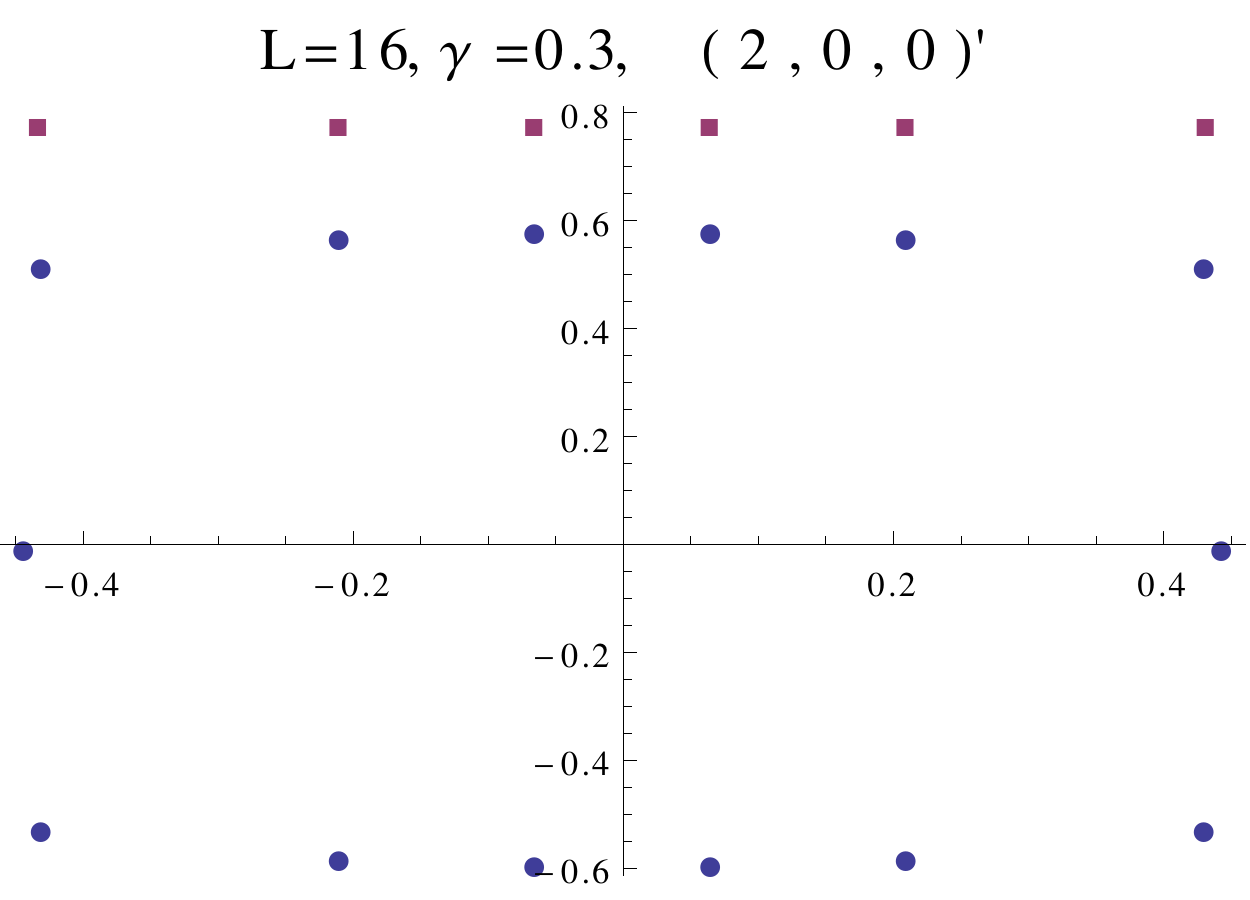}
   \\
   \includegraphics[width=90mm,height=70mm]{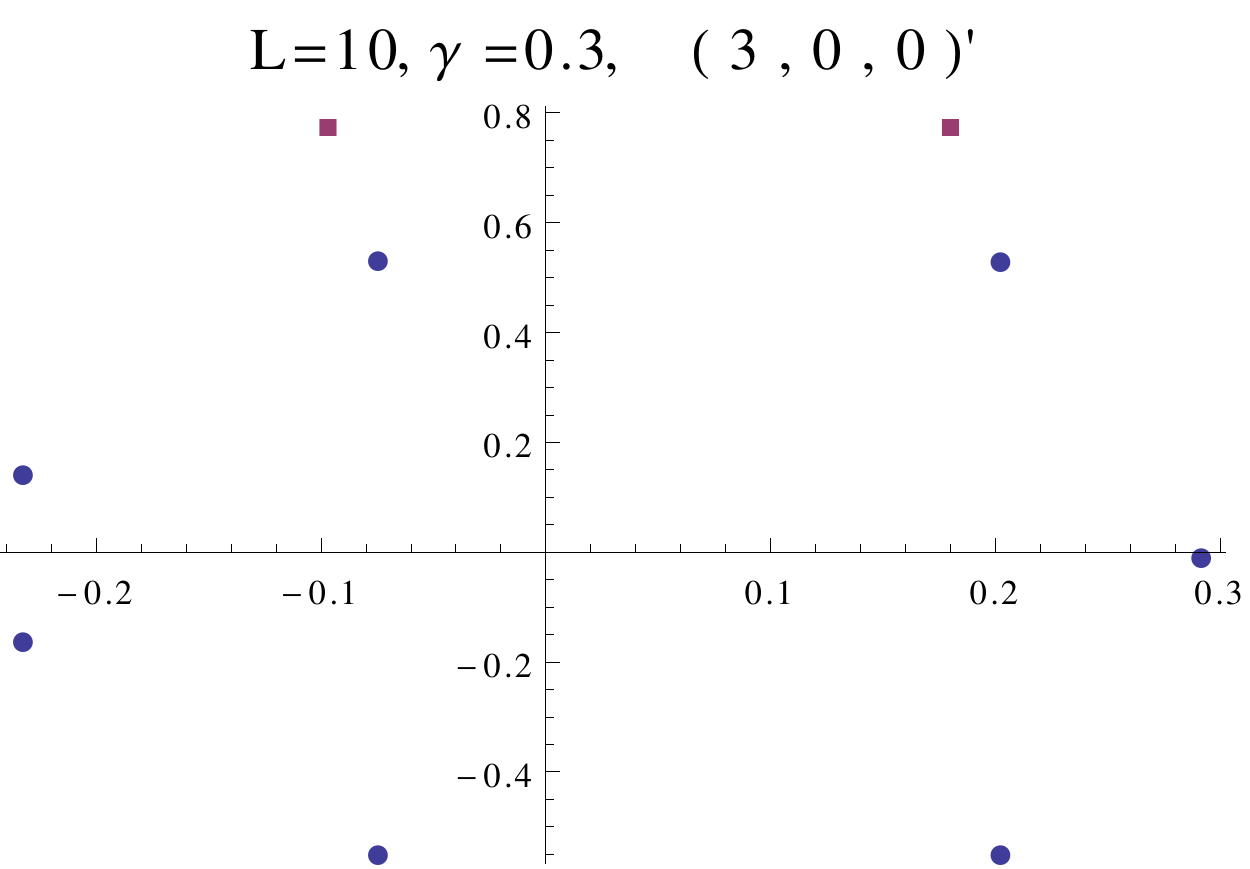}
 \end{center}
\caption{Configuration of the $\lambda$ (in blue) and $\mu$ (in purple) roots corresponding to the states $(1,0,0)'$, $(2,0,0)'$ for a system of size $L=16$, and to the state $(3,0,0)'$ for a system a size $L=10$, at $\gamma=\frac{3}{10}$ and zero twist. These configurations are deformed as the twist angles are increased, but do not seem to undergo any qualitative change. The configuration corresponding to $(3,0,0)'$ is made of the usual sea of ${L \over 2}-3$ 2-strings of $\lambda$-roots and ${L \over 2}-3$ $\mu$-roots with imaginary part $\pi \over 4$, as well as one real $\lambda$-root and one 2*-string of $\lambda$-roots, with imaginary parts close to $\pm {\gamma \over 2}$.
}                                 
\label{fig:En00primeg030L16}                          
\end{figure}

\end{itemize}

\section{Twisted Bethe Ansatz equations for the $a_3^{(2)}$ chain}
\label{app:twistedBAE}

In this appendix, we revisit the derivation of the Bethe ansatz equations for the $a_3^{(2)}$ chain \cite{GalleasMartins04} by generalizing it to arbitrary values of the twist parameters $\phi_1$ and $\phi_2$. We also found these equations in \cite{MartinsNienhuis}, without a derivation. 

Consider therefore the $a_3^{(2)}$ transfer matrix $T^{(L)}(\lambda)$ (\ref{T6GM}) acting on a system of horizontal size $L$, with a general twist 
\begin{equation}
 \tau_{\mathcal{A}} = \mathrm{e}^{-2\mathrm{i}(\phi_1 s_1^{(z)}+\phi_2 s_2^{(z)})} \,.
\end{equation}
We follow closely the notations of \cite{GalleasMartins04}, except for the relabeling of the indices introduced in section \ref{section:twocolourstructure}.

First it is useful to introduce 
\begin{equation}
 q_{-2}, q_{-1}, q_{1}, q_{2} = \mathrm{e}^{\mathrm{i} \left(\phi_1+ \phi_2  \right)} ,  \mathrm{e}^{\mathrm{i} \left(\phi_1- \phi_2  \right)} ,  \mathrm{e}^{\mathrm{i} \left(-\phi_1+ \phi_2  \right)} ,  \mathrm{e}^{\mathrm{i} \left(-\phi_1- \phi_2  \right)} 
\end{equation}
The transfer matrix (\ref{T6GM}) is therefore rewritten 
\begin{equation}
 T^{(L)} = \sum_{\alpha = -2,-1,1,2} \mathcal{T}^{(L)}_{\alpha, \alpha} q_{\alpha} \,,
 \label{T:twistedtrace}
\end{equation}
where $\mathcal{T}^{(L)} = R_{\mathcal{A}1}\ldots R_{\mathcal{A}L}$ is the so-called monodromy matrix, independent of the twist and thereby identical to that of \cite{GalleasMartins04}. Namely, its matrix action on the auxilliary space can be written as 
\begin{equation}
 \mathcal{T}^{(L)}(\lambda) 
=
\begin{blockarray}{ccccc}
 & \downarrow\downarrow & \downarrow\uparrow & \uparrow\downarrow & \uparrow\uparrow  \\
  \begin{block}{c(cccc)}
\downarrow\downarrow &    B(\lambda) 	  & B_{-1}(\lambda) 	& B_1(\lambda) 		& F(\lambda) \\  
\downarrow\uparrow &   C_{-1}(\lambda) & A_{-1,-1}(\lambda) 	& A_{-1,1}(\lambda) 	& B_{-1}^*(\lambda) \\
\uparrow\downarrow &   C_1(\lambda) & A_{1,-1}(\lambda) 	& A_{1,1}(\lambda)	        & B_1^*(\lambda) \\
\uparrow\uparrow &    C(\lambda) 	  & C_{-1}^*(\lambda)  & C_1^*(\lambda)    & D(\lambda) \\
  \end{block}
\end{blockarray}
 \,,
\end{equation}
where we also have indicated the decomposition in terms of the spins $S_1^{(z)}$ and $S_2^{(z)}$. All entries of $\mathcal{T}^{(L)}(\lambda)$ act as operators on the quantum space $\mathcal{V}^{\otimes L} = \left( \mathbb{C}^4\right)^{\otimes L}$, and it is easily seen, by conservation of the spins $S_1^{(z)}$ and $S_2^{(z)}$ under the action of each factor $ R_{\mathcal{A}i}$, that each non-diagonal entry of $\mathcal{T}^{(L)}(\lambda)$ corresponding to a change of either one of $S_1^{(z)}$ and $S_2^{(z)}$ on the auxilliary space must correspond to the opposite change in the quantum space, hence the entries in the upper-triangular half of $\mathcal{T}^{(L)}(\lambda)$ act as spin-raising operators on the quantum space, while those in the lower-triangular half act as spin-lowering operators.  

The problem of finding the eigenvalues of $T^{(L)}(\lambda)$, namely $ T^{(L)}(\lambda) |\phi \rangle  = \Lambda^{(4)}(\lambda) |\phi \rangle $, is rewritten as
\begin{equation}
 \left[ q_{-2} B(\lambda) + \sum_{a = \pm 1} q_a A_{a a}(\lambda) +q_{2}D(\lambda) \right] |\phi \rangle  = \Lambda^{(4)}(\lambda) |\phi \rangle \,.
\end{equation}
Consider as a {\it pseudovacuum} the reference state $\left| \Phi_{0} \right\rangle \equiv |-2 \rangle ^{\otimes L} = |\downarrow\downarrow \rangle ^{\otimes L}$. 
According to the above discussion it is annihilated by all the lower-triangular entries of $\mathcal{T}^{(L)}(\lambda)$, moreover we find for the diagonal entries 
\begin{equation} 
B(\lambda)\left| \Phi_{0} \right\rangle = \left[ a(\lambda) \right]^L \left| \Phi_{0} \right\rangle \,, \quad
A_{a,a}(\lambda)\left| \Phi_{0} \right\rangle = \left[ b(\lambda) \right]^L \left| \Phi_{0} \right\rangle \,(a=\pm 1)\,, \quad
D(\lambda)\left| \Phi_{0} \right\rangle = \left[ d_{2,2}(\lambda) \right]^L \left| \Phi_{0} \right\rangle \,, \quad
\end{equation}
implying that $\left| \Phi_{0} \right\rangle$ is an eigenvector of $T^{(L)}(\lambda)$ with eigenvalue 
\begin{equation}
 \Lambda_0^{(4)} = q_{-2} \left[ a(\lambda) \right]^L + \sum_{a = \pm 1} q_a \left[ b(\lambda) \right]^L +q_{2}\left[ d_{2,2}(\lambda) \right]^L \,.
\end{equation}
The algebraic Bethe ansatz consists in looking for eigenstates as combinations of spin-raising ({\it creation}) operators acting on this pseudovacuum. 
Consider a {\it $m_1$-particle state} $\left| \Phi_{m_1} \right\rangle$ obtained by flipping the total amount of $m_1$ spins, written as a combination of monomials of the form $\left(B_{-1}\right)^{n_{-1}}\left(B_1\right)^{n_1} F^{n_2}$, with $n_{-1} + n_1 + 2 n_2 = m_1$. We temporarily forget about the creation operator $F$, such that the remaining terms can be written as a product 
\begin{equation}
\prod_{j=1,\ldots,m_1}\left( \mathcal{F}_j^- B_{-1}(\lambda_j) + \mathcal{F}_j^+ B_{1}(\lambda_j) \right) \,,
\end{equation}
parametrized by a set of $m_1$ rapidities $\lambda_1, \ldots, \lambda_{m_1}$, the so-called {\it Bethe roots}. 

It turns out that such a parametrization still exists when allowing back operators $F$, namely 
\begin{equation}
 \left| \Phi_{m_1} \right\rangle = \vec{\Phi}_{m_1}(\lambda_1, \ldots, \lambda_{m_1}) \cdot \vec{\mathcal{F}}\left| \Phi_{0} \right\rangle   
 =  \Phi^{a_1,\dots,a_{m_1}}_{m_1}(\lambda_1, \ldots, \lambda_{m_1}) \mathcal{F}_{a_1,\dots,a_{m_1}}\left| \Phi_{0} \right\rangle     \,,
 \label{NestedBAEphi1}
\end{equation}
where $\vec{\mathcal{F}}$ is a $2^{m_1}$-dimensional complex vector, and $\vec{\Phi}_{m_1}(\lambda_1, \ldots, \lambda_{m_1})$ is a $2^{m_1}$-dimensional vector whose components are built from the creation operators $B_1, B_{-1}, F$, with coefficients expressed as functions of the $m_1$ rapidities ({\it Bethe roots}) $\lambda_1, \ldots, \lambda_{m_1}$.

The action of $ \left[ q_{-2} B(\lambda) + \sum_{a = \pm 1} q_a A_{a a}(\lambda) +q_{2}D(\lambda) \right]$ on $ \left| \Phi_{m_1} \right\rangle$ can be computed using the commutation relations between the creation operators  $B_1, B_{-1}, F$ and the diagonal operators $B, A_{a,a}, D$. These can be obtained from the ``RTT relation" 
\begin{equation}
 R_{\mathcal{A}_1\mathcal{A}_2}(\lambda-\mu) \mathcal{T}_{\mathcal{A}_1}(\lambda) \otimes \mathcal{T}_{\mathcal{A}_2}(\mu) =   \mathcal{T}_{\mathcal{A}_2}(\mu) \otimes \mathcal{T}_{\mathcal{A}_1}(\lambda)  R_{\mathcal{A}_1 \mathcal{A}_2}(\lambda-\mu) \,,
\end{equation}
and are given in \cite{GalleasMartins04}. After some calculation, we find that the action of $T^{(L)}(\lambda)$ writes as the sum of one term proportional to $\left| \Phi_{m_1} \right\rangle$, and two kinds of unwanted terms, which must be cancelled to ensure that $\left| \Phi_{m_1} \right\rangle$ is an eigenvector of $T^{(L)}(\lambda)$ : 
\begin{itemize}
 \item The cancellation of the first kind of unwanted terms (``easy" unwanted terms) is ensured by imposing a recursion relation between vectors $\vec{\Phi}_{m_1}(\lambda_1, \ldots, \lambda_{m_1})$ for successive values of $m_1$. This recursion relation can be written under the symbolic form
\begin{eqnarray}
\vec{\Phi}_{m_1}(\lambda_1, \ldots, \lambda_{m_1}) &=& \vec{B}(\lambda_1) \vec{\Phi}_{m_1-1}(\lambda_2, \ldots, \lambda_{m_1})  \nonumber \\
+ \sum_{j=2}^{m_1} F(\lambda_1)&  \vec{\xi}(\lambda_1-\lambda_j)&\vec{\Phi}_{m_1-2}(\lambda_2, \ldots,\lambda_{j-1},\lambda_{j+1},\ldots, \lambda_{m_1}) B(\lambda_j)  \,,
\label{eq:BAE:recursion}
\end{eqnarray}
where $\vec{B}(\lambda)$ is the 2-dimensional vector of elements $B_{\pm 1}(\lambda)$, $\vec{\xi}(\lambda)$ is a 4-dimensional vector, and we have ommited some scalar coefficients for clarity. 
 
 \item The cancellation of the second kind of unwanted terms (``standard" unwanted terms)  amounts to a constraint on the vector $\vec{\mathcal{F}}$. Namely, $\vec{\mathcal{F}}$ is required to be a solution for the following {\it auxilliary} eigenvalue problem
 \begin{equation}
 T^{(2)}(\lambda,\{\lambda_j\}) \vec{\mathcal{F}} = \Lambda^{(2)}(\lambda,\{\lambda_j\}) \vec{\mathcal{F}} \,.
 \end{equation}
$ T^{(2)}(\lambda,\{\lambda_j\})$ is the (twisted) {inhomogeneous auxilliary transfer matrix} acting on $(\mathbb{C}^2)^{\otimes L}=\{-1,1\}^{\otimes L}$ as 
 \begin{equation}
 T^{(2)}(\lambda,\{\lambda_j\})_{a_1, \ldots, a_{m_1}}^{b_1, \ldots, b_{m_1}} = \sum_{{\alpha_0} = \pm 1} q_{\alpha_0} {r^{(2)}}_{\alpha_0\, a_1}^{b_1 \, \alpha_1}(\lambda-\lambda_1) \ldots  {r^{(2)}}_{\alpha_{L-1}\, a_{m_1}}^{b_{m_1} \, \alpha_0}(\lambda-\lambda_{m_1}) \,,
 \end{equation}
 where the auxilliary R matrix $r^{(2)}$ acting in $\left\{ -1,1 \right\} \otimes \left\{ -1,1 \right\}$ is found to be that of integrable the six-vertex model, 
 while the eigenvalue $\Lambda^{(2)}(\lambda,\{\lambda_j\}) \vec{\mathcal{F}}$ must obey the following first set of Bethe ansatz equations 
\begin{equation} 
q_2 \left[a (\lambda_j) \over b(\lambda_j) \right]^L = \Lambda^{(2)} \left(\lambda_j , \{ \lambda_l\}\right) \, \quad (j=1,\ldots m_1) \,.
\label{eq:BAE1:raw}
\end{equation}
\end{itemize}

Given such a solution of the auxilliary problem, the eigenvalue of the transfer matrix $T^{(L)}$ is obtained as 
\begin{equation}
 \Lambda^{(4)}(\lambda) = \left[a(\lambda)\right]^L q_2 \prod_{i=1}^{m_1}\frac{a\left( \lambda_i - \lambda \right)}{b\left( \lambda_i - \lambda \right)}
 + \left[d_{2,2}(\lambda)\right]^L q_{-2} \prod_{i=1}^{m_1}\frac{b\left( \lambda_i - \lambda \right)}{d_{2,2}\left( \lambda_i - \lambda \right)} 
 + \left[b(\lambda)\right]^L \Lambda^{(2)}(\lambda, \{ \lambda_l\}) \prod_{i=1}^{m_1}\frac{1}{b\left( \lambda_i - \lambda \right)} \,.
\end{equation}

The last step is now to solve the auxilliary problem, which is simply that of a twisted integrable six-vertex model with inhomogeneous spectral parameters. Once again a reference state can be taken as $\vec{\mathcal{F}} = \{-1\}^{\otimes m_1}$, which corresponds to a state  $ \left| \Phi_{m_1} \right\rangle$ where $m_1$ spins of the first colour have been raised, namely $S_1^{(z)} = -\frac{L}{2}+m_1$, $S_2^{(z)} = -\frac{L}{2}$. From this reference state excitations can be created by flipping a given number $m_2$ of components of $\vec{\mathcal{F}}$. Once again these are parametrized by a set of $m_2$ auxilliary Bethe roots $\mu_1, \ldots, \mu_{m_2}$, solution of the following set of Bethe equations 
\begin{equation} 
(q_{-1})^2 \prod_{j=1}^{m_1} \frac{\sinh (2(\mu_k-\lambda_j-\mathrm{i}\frac{\gamma}{2}))}{\sinh (2(\mu_k-\lambda_j+\mathrm{i}\frac{\gamma}{2}))} =
\prod_{l=1, l\neq k}^{m_2} \frac{\sinh (2(\mu_k-\mu_l-\mathrm{i}\gamma))}{\sinh (2(\mu_k-\mu_l+\mathrm{i}\gamma))} 
 \,,
\end{equation}
while the eigenvalue $\Lambda^{(2)}(\lambda, \{ \lambda_l\})$ is obtained as  
\begin{equation}
\Lambda^{(2)}(\lambda, \{ \lambda_l\}) = q_{-1} \tilde{G}_{-1}(\lambda) + q_{1} \tilde{G}_{1}(\lambda) \,.
\end{equation}
We have used the notation $\tilde{G}_{\pm 1}(\lambda) = \left(\prod_{i=1}^{m_1} b(\lambda_i-\lambda)\right) G_{1(2)}(\lambda)$, and refer to \cite{GalleasMartins04} for the definition of $G_1$ and $G_2$. At $\lambda = \lambda_j$ only $\tilde{G}_{-1}(\lambda_j)$ is non zero, and the first set of Bethe ansatz equations (\ref{eq:BAE1:raw}) is rewritten as
\begin{equation}
\mathrm{e}^{2\mathrm{i}\phi_1}\left(\frac{\sinh (\lambda_i-\mathrm{i}\frac{\gamma}{2})}{\sinh (\lambda_i+\mathrm{i}\frac{\gamma}{2})} \right)^L = 
\prod_{j=1, j\neq i}^{m_1} \frac{\sinh (\lambda_i-\lambda_j-\mathrm{i}\gamma)}{\sinh (\lambda_i-\lambda_j+\mathrm{i}\gamma)} 
\prod_{k=1}^{m_2} \frac{\sinh (2(\lambda_i-\mu_k+\mathrm{i}\frac{\gamma}{2}))}{\sinh (2(\lambda_i-\mu_k-\mathrm{i}\frac{\gamma}{2}))} 
\end{equation}.

Let us sum up : the eigenvalues of the transfer matrix $T^{(L)}$ are parametrized by a set of $(m_1,m_2)$ Bethe roots, and the following correspondence stems from our analysis
\begin{eqnarray}
 S_1^{(z)} &=& -{L \over 2} +  m_1 - m_2  \,, \nonumber \\
 S_2^{(z)} &=& -{L \over 2} + m_2 \,,
 \label{relateSandm}
\end{eqnarray}
(note that we might as well add an extra minus sign in front of the definition of $S_1^{(z)}$ or $S_2^{(z)}$ without changing anything to the results). In particular, the ground state sector ($S_1^{(z)} = S_2^{(z)} = 0$) corresponds to $m_1 = L$ and $m_2={L \over 2}$.

The roots $\lambda_1, \ldots, \lambda_{m_1}$, $\mu_1, \ldots, \mu_{m_2}$ are solution of the twisted Bethe ansatz equations \cite{MartinsNienhuis}

\begin{eqnarray}
\mathrm{e}^{2\mathrm{i}\phi_1}\left(\frac{\sinh (\lambda_i-\mathrm{i}\frac{\gamma}{2})}{\sinh (\lambda_i+\mathrm{i}\frac{\gamma}{2})} \right)^L &=& 
\prod_{j=1, j\neq i}^{m_1} \frac{\sinh (\lambda_i-\lambda_j-\mathrm{i}\gamma)}{\sinh (\lambda_i-\lambda_j+\mathrm{i}\gamma)} 
\prod_{k=1}^{m_2} \frac{\sinh (2(\lambda_i-\mu_k+\mathrm{i}\frac{\gamma}{2}))}{\sinh (2(\lambda_i-\mu_k-\mathrm{i}\frac{\gamma}{2}))}  \nonumber \\
\mathrm{e}^{2\mathrm{i}\left(\phi_1 - \phi_2\right)} \prod_{j=1}^{m_1} \frac{\sinh (2(\mu_k-\lambda_j-\mathrm{i}\frac{\gamma}{2}))}{\sinh (2(\mu_k-\lambda_j+\mathrm{i}\frac{\gamma}{2}))} &=& 
\prod_{l=1, l\neq k}^{m_2} \frac{\sinh (2(\mu_k-\mu_l-\mathrm{i}\gamma))}{\sinh (2(\mu_k-\mu_l+\mathrm{i}\gamma))} \,,
\end{eqnarray}
and the corresponding eigenvalues of $T^{(L)}$, in the notations of \cite{GalleasMartins04} (which means in particular that we have traded back the notation $d_{2,2}(\lambda)$ for $d_{4,4}(\lambda)$), are 
\begin{eqnarray}
 \Lambda^{(4)}(\lambda) &=& \mathrm{e}^{\mathrm{i} \left( \phi_1 + \phi_2 \right)} \left[a(\lambda)\right]^L \frac{Q_1\left( \lambda + \mathrm{i}\frac{\gamma}{2} \right)}{Q_1\left( \lambda - \mathrm{i}\frac{\gamma}{2} \right)}
 +  \mathrm{e}^{\mathrm{i} \left( -\phi_1 - \phi_2 \right)} \left[d_{4,4}(\lambda)\right]^L \frac{Q_1\left( \lambda - \mathrm{i}\frac{5\gamma}{2} + \mathrm{i}\frac{\pi}{2} \right)}{Q_1\left( \lambda - \mathrm{i}\frac{3\gamma}{2} + \mathrm{i}\frac{\pi}{2} \right)}  \nonumber \\
 & & +  \left[b(\lambda)\right]^L \left( \mathrm{e}^{\mathrm{i} \left( \phi_1 - \phi_2 \right)} G_1\left(\lambda  \right)  + \mathrm{e}^{\mathrm{i} \left(- \phi_1 + \phi_2 \right)} G_2\left(\lambda  \right) \right) \,,
\end{eqnarray}
where $Q_1(\lambda) \equiv \prod_{i=1}^{m_1} \sinh\left(\lambda - \lambda_i\right)$ and $Q_2(\lambda) \equiv \prod_{i=1}^{m_2} \sinh\left(\lambda - \mu_i\right)$.

These are the results we have been using in the main text.

\end{document}